\documentclass[12pt]{article}
\usepackage{graphicx,amsmath,amssymb}
\usepackage{subcaption}
\usepackage{lineno}
\usepackage{units}
\usepackage{mwe}

\newlength{\dinwidth}
\newlength{\dinmargin}
\renewcommand{\vec}[1]{\boldsymbol{#1}}

\def\lapproxeq{\lower .7ex\hbox{$\;\stackrel{\textstyle                                                    
<}{\sim}\;$}}                                                    
\def\gapproxeq{\lower .7ex\hbox{$\;\stackrel{\textstyle                                                    
>}{\sim}\;$}}                                                    
\def\be{\begin{equation}}                                                    
\def\ee{\end{equation}}                                                    
\def\bea{\begin{eqnarray}}                                                    
\def\eea{\end{eqnarray}}

\def\sh{\hat s}
\def\sh2{{\hat s}^2}

\begin{document}
\titlepage                                                    
\begin{flushright}                              
HIP-2022-14/TH \\                       
IPPP/22/37  \\  
LTH 1305 \\                             
\vspace{0.3cm}                 
\today \\                                                    
\end{flushright} 
\vspace*{0.5cm}
\begin{center}

{\Large \bf  Exclusive $J/\psi$ and $\Upsilon$ production \vspace{0.3cm}
in high energy $pp$ and $p$Pb collisions}\\

\vspace*{1cm}
                                                   
C.~A.~Flett$^{a,b}$, S.~P.~Jones$^c$, A.~D.~Martin$^c$, M.~G.~Ryskin and T.~Teubner$^d$\\                                                    
                                                   
\vspace*{0.5cm}  
\fontsize{10.47}{1}
$^a${\it Department of Physics, University of Jyv\"{a}skyl\"{a}, P.O. Box 35, 40014 University of Jyv\"{a}skyl\"{a}, Finland}\\
$^b${\it Helsinki Institute of Physics, P.O. Box 64, 00014 University of Helsinki, Finland}\\
$^c${\it Institute for Particle Physics Phenomenology, Durham University, Durham, DH1 3LE, U.K.} \\                              
$^d${\it Department of Mathematical Sciences, University of Liverpool, Liverpool, L69 3BX, U.K.}\\

\vspace*{1cm}                                                    
                                                    
\begin{abstract}
We present a formalism for determining the cross section for exclusive heavy vector meson production ($J/\psi, \Upsilon$) as a function of rapidity, in both high energy proton-proton and proton-heavy ion collisions, at next-to-leading order in QCD. We compare and contrast the production in $pp$ and $p$Pb collisions and show how data for these processes can give information on the low $x$ gluon distribution of the proton and heavy ions at a range of different scales. 
\vspace*{0.2cm}  
 
\end{abstract}

\end{center}

\section{Introduction}
\begin{figure} [t]
\centering
\includegraphics[scale=0.7]{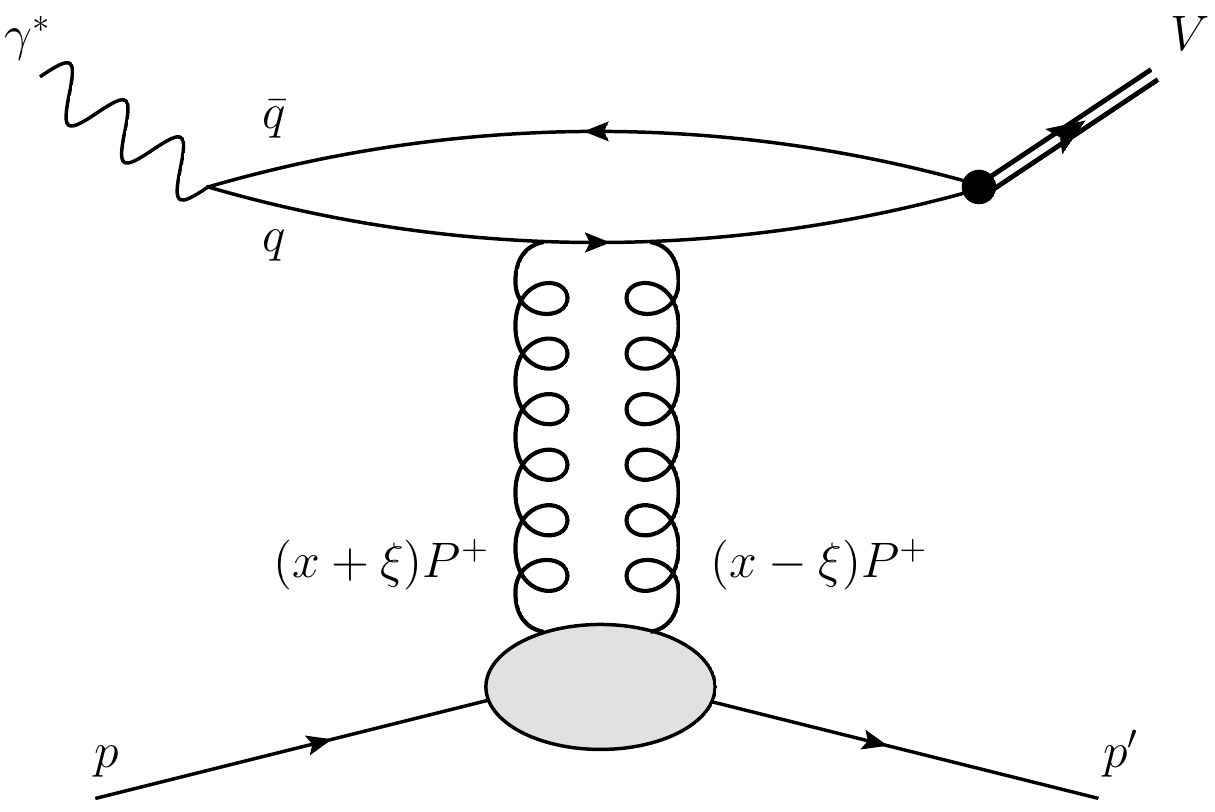}
\begin{center}
\caption{\sf{A schematic diagram of high-energy exclusive vector meson production, $\gamma^* +
p\to V+p$. The factorized form follows since, in the
proton rest frame, the formation time $\tau_f\simeq  2E_\gamma / (Q^2+M^2_{V})$ is much greater than the $q\bar q$-proton interaction time. $Q$ is
the virtuality of the photon and the lightcone momentum $P^+ = (p+p')/2$. For the photoproduction set-up considered here, we have $Q^2 = 0$. }}
\label{f1-l}
\end{center}
\end{figure}

There is a long history of experimental and theoretical study of exclusive vector meson production in high-energy proton-proton collisions. In particular, data for the differential cross section
$\text{d}\sigma (p+p\to p+V+p)/\text{d}Y$
has come under theoretical scrutiny for vector mesons $V=J/\psi, \Upsilon(1S)$, where the $+$ signs denote large rapidity gaps between the rapidity $Y$ of the vector meson and the outgoing protons, ensuring the exclusivity of the experimental measurements. The theoretical description proceeds by first calculating the differential cross section for the exclusive photoproduction subprocess $\gamma + p\to V+p$ for which the dominant QCD diagram is sketched in simplified form in Fig.~\ref{f1-l}. At high energy the process is driven by the behaviour of the gluon parton distribution of the proton at small momentum fraction.  
In this work, we apply the theoretical framework, based on collinear factorisation and developed to NLO in perturbative QCD in~\cite{Ivanov:2004vd, Jones:2016ldq}, to $p$Pb collisions. Other approaches include, for example, models based on the Colour-Glass-Condensate~(CGC)~\cite{Iancu:2003ge, Kowalski:2006hc} and $k_T$-factorisation frameworks~\cite{Jones:2016icr}, as well as LO perturbative QCD~\cite{Guzey:2013taa}. Note that the two exchanged gluons in Fig.~\ref{f1-l} carry different fractions of the incoming proton momentum, so we are dealing with a generalised, skewed gluon distribution. The net momentum fraction transferred from the proton to the vector meson is $2\xi$, thus for this exclusive process $2\xi$ plays the role of the usual variable `$x$' in DIS. 
For our high energy process we have $x-\xi \ll x+\xi \ll 1$ and the cross section can be accurately estimated in terms of the conventional gluon distribution $g(2\xi)$ and a skewing correction~\cite{Shuvaev:1999ce, Shuvaev:1999fm}. 
Recall that a detailed NLO analysis of the current inclusive data determines the NLO gluon distribution down to $2\xi \simeq 10^{-4}$~\cite{Bailey:2020ooq}. For smaller momentum fractions the uncertainties become too large, see also~\cite{Flett:2019pux}. However, the exclusive $J/\psi$ data from LHC~\cite{LHCb:2014acg,LHCb:2018rcm} are found to determine the gluon distribution down to  $2\xi \sim 3 \times 10^{-6}$. 

In Section 2 we note that in going from the cross section for photoproduction, $\gamma + p\to V+p$, to that for $p+p\to p+V+p$, we need to allow for photon emissions from both incoming protons. Moreover, we have to evaluate the photon flux and the survival factors of the rapidity gaps, i.e. the probabilities that the rapidity gaps do not get populated by additional emissions.
Section 3 concerns exclusive vector meson production in $p$Pb collisions. Data for such a process appear to have the advantage of having a much larger cross section; enhanced by the large charge of the heavy Pb ion, that is by a factor $Z^2=6724$. However, when we come to evaluate the cross section we find the theoretical formalism is much more complicated than that for $pp$ collisions.
In Section 4 we compare the differential cross section data from ALICE and CMS at the proton-nucleon collision energy $\sqrt{s_{pN}} = 5.02~$TeV for the process $p+{\rm Pb} \to p+V +{\rm Pb}$, where $V = J/\psi, \Upsilon$,  with our corresponding theoretical results. We then provide predictions at $\sqrt{s_{pN}} = 8.16~$TeV, before presenting our conclusions and outlook in Section 5.

\section{Exclusive $p+p\to p+V+p$ production}
\begin{figure} [t]
\centering
\includegraphics[width=0.4\textwidth]{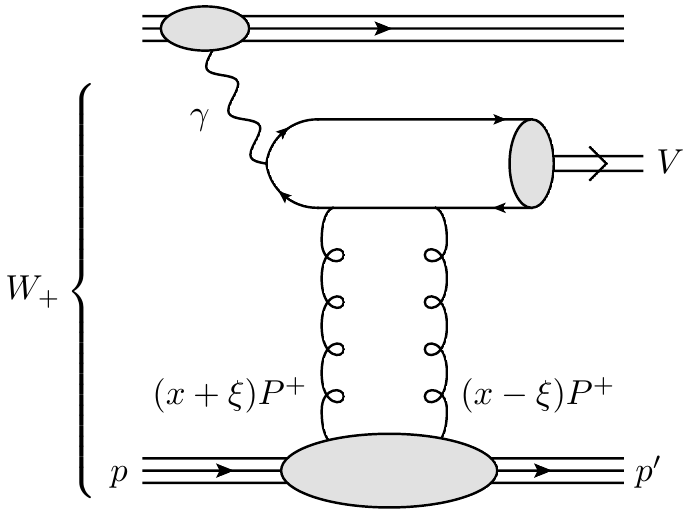}
\qquad
\includegraphics[width=0.4\textwidth]{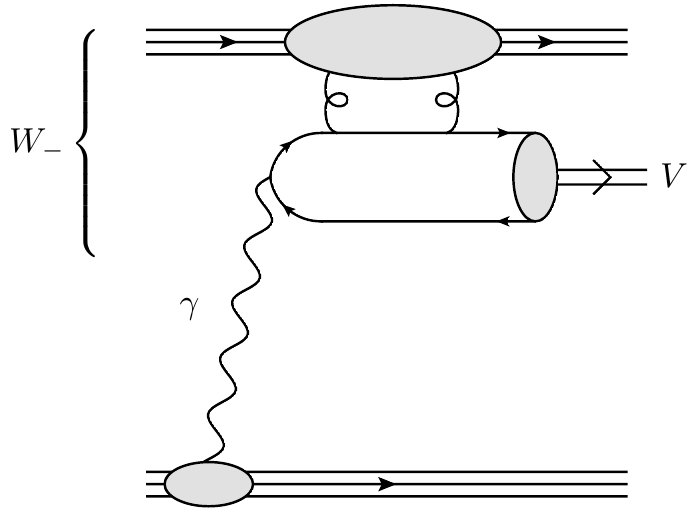}
\caption{\sf{The two diagrams describing exclusive heavy vector meson production in $p+p\to p+V+p$, at the LHC. The $W_+$
and $W_-$ contributions arise in the ultraperipheral description of the $\gamma+p \to V + p$ subprocess, see the
text for details. In the $p$Pb mode, either the upper or lower proton is replaced by a Pb-ion.
} }
\label{wpwm}
\end{figure}

We begin with $pp$ collisions, and leave the discussion of $p$Pb collisions until the next section. The procedure for calculating the cross section for the high-energy exclusive process $p+p\to p+V+p$ (where the vector meson $V=J/\psi$ or $\Upsilon$ and where the + signs represent rapidity gaps) is described in $\cite{Jones:2016icr, Flett:2021fvo}$. As mentioned above, we first calculate the cross section for the exclusive photoproduction process
\be
\label{A1}
\sigma_W(\gamma p)~\equiv~\sigma(\gamma+p \to V+p)
\ee
where 
$W$ is the $\gamma p$ centre-of-mass energy.
This process is driven by the two amplitudes sketched in Fig.~\ref{wpwm}.  In  the left diagram the photon is radiated by the upper proton and then the vector meson, $V$, is created on the lower proton via the $\gamma+p\to V+p$ reaction. The diagram on the right shows the other possibility where the photon is emitted from the lower proton while the upper proton acts as the target.

When the meson $V$ is detected in the forward region, the corresponding $\gamma p$ energies are quite different
\begin{equation}
\label{W}
W^2_+=M_V\sqrt s~e^Y\;\;\;\mbox{and}\;\;\; W^2_-=M_V\sqrt s~e^{-Y},
\end{equation}
where $Y$ is the rapidity of the vector meson in the laboratory (i.e. $pp$ centre-of-momentum) frame. 
Since the $\gamma p\to V+p$  cross section steeply grows with energy, the dominant contribution comes from the first amplitude which corresponds to a larger energy $W_+$.

The cross section for exclusive $V$ production in ultraperipheral $pp$ collisions is therefore given by the sum\footnote{The interference between the two amplitudes of the two subprocesses is small
since the transverse momentum of the proton which radiates the photon is much smaller
than that of the target proton. Thus we may neglect the interference term at the accuracy
we are aiming at here.} of the exclusive photoproduction cross sections, $\sigma_\pm (\gamma p)$ of eqn.~\eqref{A1}, at the two $\gamma p$ energies $W_\pm$
\be
\label{sig-t} 
\frac{d\sigma(p+p \to p + V + p)}{dY}~=~S^2(W_+) \left( k_+ \frac{dn_p }{dk_+} \right)\sigma_+ (\gamma p)~+~S^2(W_-) \left( k_- \frac{dn_p }{dk_-} \right)\sigma_- (\gamma p).
\ee
Note that the subprocess cross sections are weighted by the survival factors $S^2(W_\pm)$ which account for the probability that the rapidity gap between the vector meson and the target proton is not populated by soft interactions which would destroy the exclusivity of the event; and by the photon fluxes $dn_p /dk_\pm$ for photons of energy $k_\pm=x_\pm \sqrt{s}/2$, where $x_\pm$ are the fractions of the parent proton energy carried by the photon.

The flux of photons emitted from a proton is well known~\cite{Budnev:1975poe}
\be
\label{flux0}
\frac{dn_p (x)}{dx}=\frac{\alpha^{\rm QED}}{\pi^2 x}\int\frac{d^2q_\perp 
}{q^2_\perp+x^2m^2_p}\left(\frac{q^2_\perp}{q^2_\perp+x^2m^2_p} (1-x)F_E(Q^2)+\frac{x^2}2F_M(Q^2)\right),
\ee
where $q_\perp$ is the photon transverse momentum, $m_p$ is the proton mass and $F_{E,M}$ are the proton form factors
\begin{align}
F_E(Q^2) &= (4m^2_pG^2_E(Q^2)+Q^2G_M(Q^2))/(4m^2_p+Q^2)\,, \\[10pt]
F_M(Q^2) &= G_M^2(Q^2)\,.
\end{align}
The scale  $Q^2=(q^2_\perp+x^2m^2_p)/(1-x)$, while $G_E$  and $G_M$ are the  ‘Sachs’ form factors, which may be expressed  
in dipole forms 
\begin{align}
G_E(Q^2) &= 1/(1+Q^2/0.71\mbox{GeV}^2)^2,  \\[10pt]
G_M(Q^2) &= 2.79/(1+Q^2/0.71\mbox{GeV}^2)^2.
\end{align}

The rapidity gap survival factors, $S^2(W_\pm )$,
are calculated in impact parameter, $b$, space using
\be
\label{op}
S^2(W_\pm,b))~=~\exp(-\Omega(b,W_\pm))\,,
\ee
where $\Omega(b,W)$ is the opacity (i.e. optical density) of the proton-proton interaction at the energy $W$ and the vector valued impact parameter $\vec{b} = (b_x, b_y)$, see e.g.~\cite{Khoze:2017sdd}.

The results as a function of the rapidity of the vector meson are tabulated for different collider energies in \cite{Jones:2016icr} and \cite{Flett:2021fvo} for $V=J/\psi$ and $\Upsilon$, respectively.

\section{Exclusive $p$ + Pb $\to p+V+$ Pb production}
 In a proton-lead collision the photon flux radiated coherently by the lead ion is strongly enhanced by the factor $Z^2=6724$. So, as mentioned before, at first sight the  
 amplitude with the photon emitted by the lead ion (say, Fig.~\ref{wpwm} left) should dominate. However the situation is not so simple.

\subsection{The different contributions}

Firstly, in the $\gamma$+Pb$\,\,\rightarrow\,V+$Pb process, the nucleons situated at the same impact parameter (i.e. on the line directed along the beam) may interact coherently as well. That is we have a competition between the factor $Z^2=6724$ and a coherent factor of about $A^{4/3}= 1232$, see~\cite{Harland-Lang:2018ytk} for more details. Thus the proton induced rate is suppressed only by a factor of 5 relative to that induced by Pb.

Moreover, as it is seen from the right hand side of (\ref{flux0}), we have essentially a logarithmic integral $\int dq^2_\perp/q^2_\perp$ in the interval from $q_\perp \simeq xm_p$ up to the value $\sim 1/R$ limited by the form factors. That is, within the leading-logarithmic approximation,  
the photon flux is proportional to $\ln(1/(x\,m_pR))$, where
$x_\pm=(M_V/\sqrt s)e ^{\pm Y}$ is the proton momentum fraction carried by the photon and $R$ is the radius of the photon emitter. Thus, for a large rapidity $Y$, the flux radiated by the proton may be enhanced by the ratio
\be
\ln(1/(x_- m_p R_p))~/~\ln(1/(x_+ m_pR_{\rm Pb}))\,,
\ee 
which for an $\Upsilon$ at $Y=2.5$ reaches 6.8 at $\sqrt s=14$ TeV.
 Here $(xm_p)^2$ is the minimal  virtuality squared, $|t_{\rm min}|$, of the photon which carries the fraction $x$ of the momentum of the parent proton.  The crucial point is that to produce a heavy  $\Upsilon$ at large $Y$ the photon must carry a large fraction of the nucleon beam momentum and the value of $xm_p$ becomes comparable to the inverse ion radius, $1/R_{\rm Pb}$. That is the logarithm in the denominator of the ratio becomes small.

Besides this, the amplitude where the photon was radiated by the proton may be enhanced due to the energy dependence of the elementary photoproduction amplitude. 

Therefore, it is not evident that it is sufficient to consider only the photons radiated by the heavy ion. In particular, for the case of forward exclusive $\Upsilon$ production in the lead ion direction, the contribution caused by the photon radiated off the proton beam may even dominate at large $Y$. In our computations we will keep both amplitudes.  The kinematic configurations are illustrated~(see Fig.~\ref{configs}) and discussed in Appendix B.

\subsection{Including the probability of rapidity gap survival}
Here we recall the structure of the calculations needed to account for the gap survival probability in {\em exclusive} vector meson production in $p$Pb collisions. We will follow Section 6 of~\cite{Harland-Lang:2018ytk}, see also~\cite{Harland-Lang:2018iur}.

\begin{figure} [t]
\centering
\includegraphics[scale=0.4]{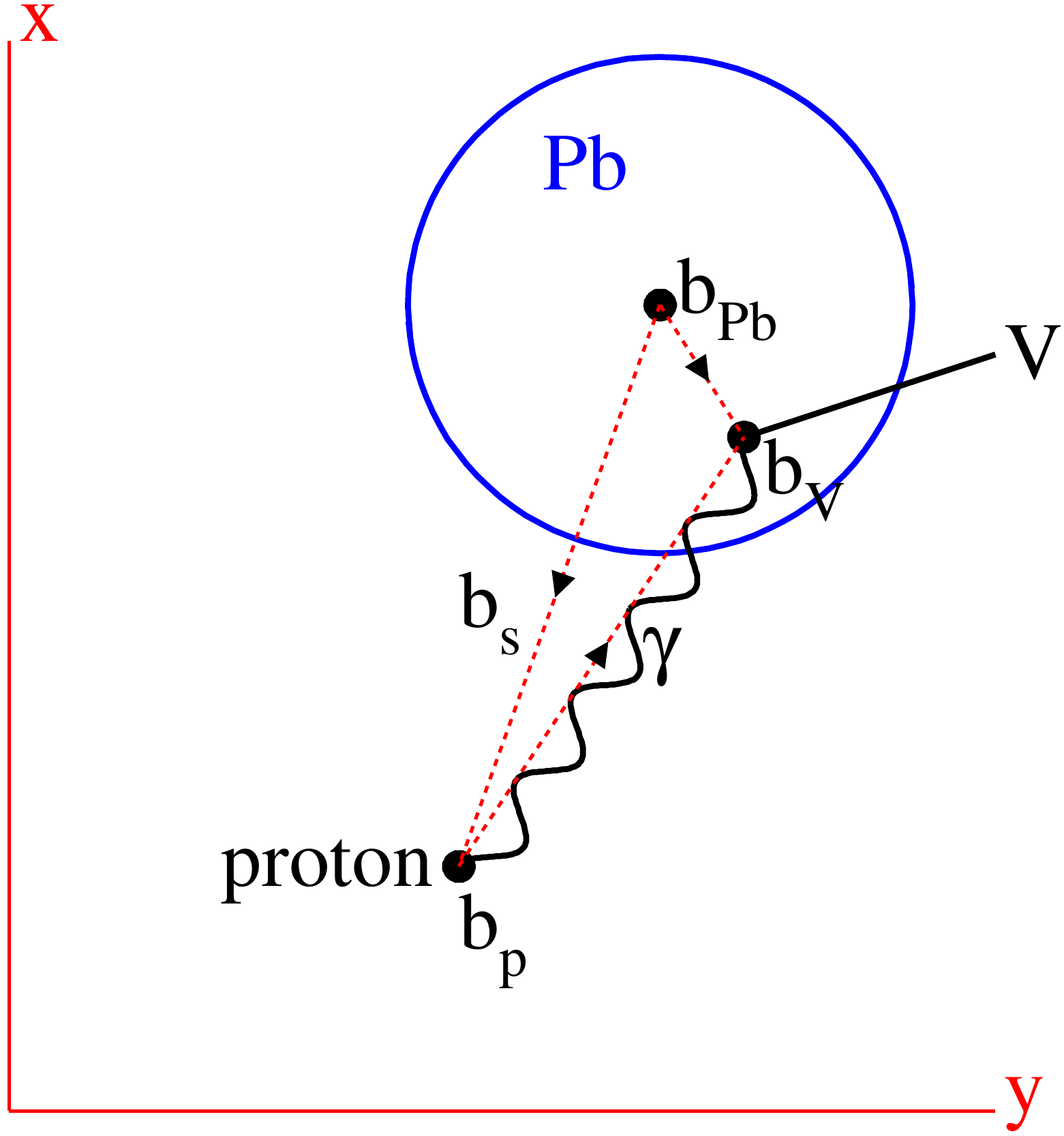}
\qquad
\includegraphics[scale=0.4]{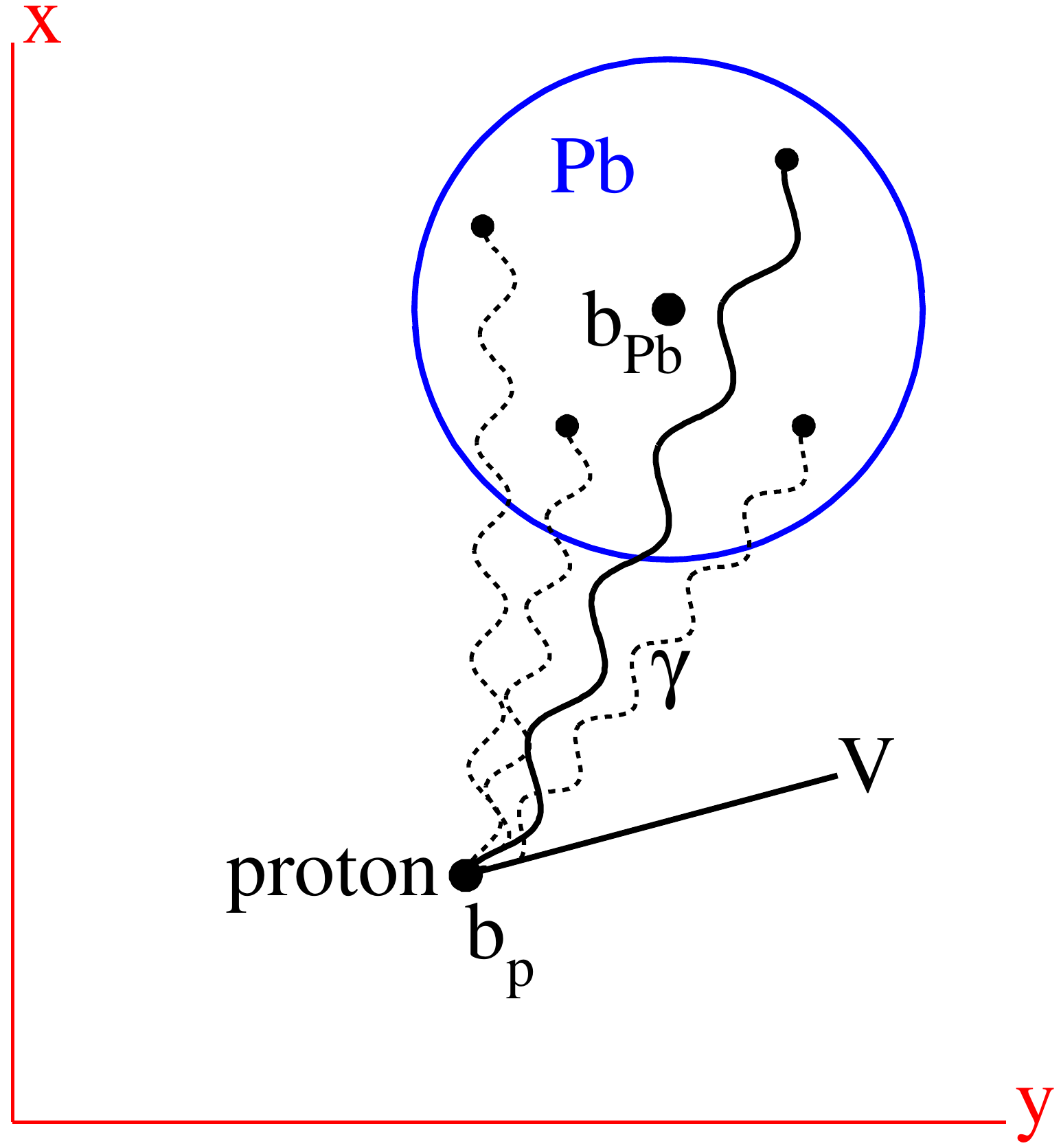}
\caption{\sf{The geometry of the $p+{\rm Pb}\to p+V+{\rm Pb}$ process in the transverse ($x,y$) plane. In the left diagram the photon is radiated off the incoming proton, whereas in the right diagram the
photons are radiated by the lead ion. The black dotted curves indicate that we add coherently the contributions of each proton inside the lead ion. Note that the photons can be emitted and absorbed at different points along the $z$ beam axis.}}
\label{f2}
\end{figure}

It is convenient to work in terms of the transverse coordinate; that is, to work in the two-dimensional impact parameter $b$ space. Let us start with the incoherent production of the vector meson in proton-ion collisions in which the photon is emitted from the incoming proton. The situation is sketched in Fig.~\ref{f2}(a). The figure shows the impact factors $\vec{b}_{\rm Pb}$, $\vec{b}_p$ and $\vec{b}_V$ corresponding respectively to the centre of the Pb ion, the proton and the produced vector meson.  Note that we show the extent in $b$ space of the Pb ion.
It is convenient to take the vertex of the meson production to be the origin of the transverse plane, that is to take $\vec{b}_V =\vec{0}$.  

\begin{figure} [t]
\centering
\includegraphics[scale=0.4]{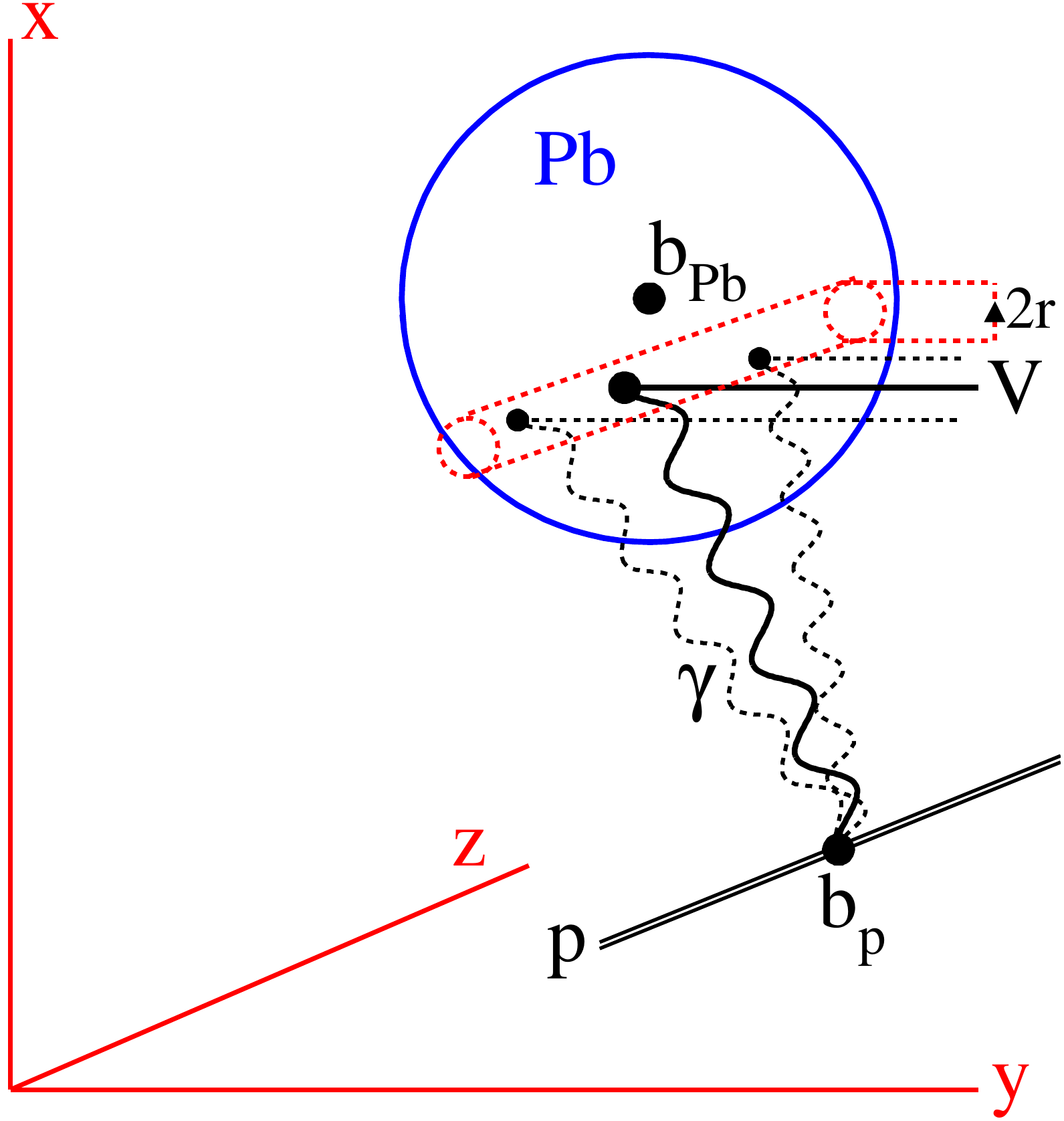}
\caption{\sf{The geometry of the $p+{\rm Pb}\to p+V+{\rm Pb}$ process, where ($x,y$) is the transverse plane and $z$ is in the direction of the beam. The black dotted curves indicate that we add coherently the contributions of each nucleon absorbing a photon inside the lead ion tube of the radius $r=\sqrt{4B_V}$ directed along the beam ($z$) axis. The transverse size of tube, $r$, is driven by the $t$-slope (i.e. the size) of the $\gamma+N\to V+N$ amplitude. The extra factor $4\pi B_V T(b_{\text{Pb}})$ in (\ref{sig-coh}) in comparison with (\ref{sig-inc}) is the number of nucleons  which may act coherently inside this tube
of transverse area $4\pi B_V$. Note that the photons can be emitted and absorbed at different points along the $z$ beam axis.}} 
\label{f3}
\end{figure}

Since we are looking for the cross section ($\sigma \propto \mathcal A^* \mathcal A)$ integrated over the momentum transferred $t$, the
 values of $b$ are the same in the $\mathcal A^*$ and $\mathcal A$ amplitudes.
Allowing for the photon flux and survival factors, we can therefore write the expression for the cross section of the {\it incoherent} $p$~+~Pb$\,\,\to\,p~+~V~$+~Pb interaction (for the case of the photon radiated by the incoming proton) as
\begin{equation}
\label{sig-inc}
\sigma_{\rm incoh}=\int d^2b_p d^2b_{\rm Pb} 
T(b_{\rm Pb})\frac{xdn_p(x,b_p)}{dx}|\mathcal{A}|^2S^2_p(b_s)
S^2_V(b_{sv})\,,
\end{equation}
where $n_p(x,b)$ is the flux of photons emitted off the proton, $\mathcal A=\mathcal A(W,b_a)$ is the $\gamma+p\to V+p$ amplitude in $b$ representation and $T(b_{\rm Pb})$ is the optical density of the Pb ion, evaluated as given in the Appendix A. Equation~\eqref{sig-inc} is written in the limit of a small size amplitude $\mathcal A$, i.e. the size of the amplitude given by the $t$-slope $B_V = B_0 + 4 \alpha' \ln \left(W/W_0 \right)$, see later, 
is much less than that for the heavy ion $R^2_{\rm Pb}\gg 4B_V$.\footnote{ Accounting, however, for the non-locality of $\mathcal A$ in the computation, i.e. the size of the proton, we replace the optical density $T(b)$~(\ref{T-opt}) by the convolution $T(b)=\frac 1{4\pi B_V}\int d^2 b'\ T(b')\exp[-( \vec{b}- \vec{b'})^2/(4B_V)]$. Analogous replacements (with the corresponding $B$ slopes) were used in the calculation of the survival factors $S^2$. For proton-nucleon interactions the slope $B_{\rm el}=20$ GeV$^{-2}$ is used for $\sqrt{s_{p\text{N}}} = 8.16$ TeV and $B_{\rm el}=19.1$ GeV$^{-2}$ for $\sqrt{s_{p\text{N}}} = 5.02$ TeV.}
The survival factors are
\begin{equation}
\label{gap-p}
S^2_p(b_s)=\exp\left(- T(b_s)\sigma(pN)\right)\,,
\end{equation}  
which is the probability not to fill the gap by secondary emissions produced in additional proton-lead interactions, where  $b_s = |\vec{b}_{\rm Pb} - \vec{b}_p|$,
and
\begin{equation}
\label{gap-v}
S^2_V(b_{sv})=\exp\left(-T(b_{sv})\sigma(VN)\right)\,,
\end{equation}
which is the probability not to fill the gap by secondaries produced in additional vector meson-Pb interactions, where $b_{sv}=|\vec{b}_{\rm Pb} - \vec{b}_V|=b_{\rm Pb}$ with our choice $\vec{b}_V=\vec{0}$. Finally,  
$\sigma(pN)$ (and  $\sigma(VN)$) are the cross sections of the proton-nucleon (vector meson-nucleon) interaction inside the ion. Recall that the expressions (\ref{gap-p}, \ref{gap-v}) are similar to the probability not to have an additional interaction in the target and the product $T(b)\sigma$ plays the role of the optical density of the target, that is the  opacity $\Omega$ in (\ref{op}).

If we were to neglect the survival factors (i.e. setting $S^2=1$) then the cross section would be
\be\sigma_{\rm incoh}=\frac{xdn_p(x)}{dx}\sigma(\gamma+p\to V+p)\cdot A\,,
\ee
where $A=208$ is the lead atomic number.

As mentioned above, besides the incoherent interaction, we need to consider coherent production as well. Here the situation is sketched in Fig.~\ref{f3}. The {\it coherent} cross section is of the form
\begin{equation}
\label{sig-coh}
\sigma_{\rm coh}=4\pi B_VF^2_{\rm Pb}(t_{\rm min})\int d^2b_p d^2b_{\rm Pb}  
T^2(b_{\rm Pb})\frac{xdn_p(x,b_p)}{dx}|\mathcal A|^2S^2_p(b_s)
S^2_V(b_{\rm Pb})\,,
\end{equation}
where the dimension of the extra $T$ factor is compensated by the $t$-slope of the $\gamma+p\to V+p$ amplitude\footnote{Recall that the $t$ behaviour of the photoproduction cross section $d\sigma(\gamma+p\to V+p)/dt\propto \exp(B_Vt)$ corresponds to the amplitude $\mathcal A(b)=\mathcal A(b=0)\exp[-b^2/(2B_V)]$.}.
The amplitude $\mathcal A$ is normalized to $\int d^2b|\mathcal A(b)|^2=\sigma(\gamma+p\to V+p)$.  

The form factor $F_{\rm Pb}$ in (\ref{sig-coh}) accounts for the
nucleon distribution in the lead ion. The point is that the coherence of the interaction with different nucleons should not be destroyed by the longitudinal component of the momentum transferred. This component is represented by $t_{\rm min}=-(xm_p)^2/(1-x)$. Since the value of $|t_{\rm min}|$ is small in our kinematics, here we use just the exponential parametrization $F_{\rm Pb}(t)={\rm exp}(t\langle r^2_{\rm Pb}\rangle /6)$, with $\langle r^2_{\rm Pb}\rangle $ being the mean radius squared of the lead ion.

We are particularly interested in the case when the photons are radiated by the lead ion. Then we have only coherent radiation to consider. Here
the situation is sketched in the right panel of Fig.~\ref{f2}. Now the cross section has the form
\be
\label{sig-Pb}
\sigma_{\rm Pb}=\sigma(\gamma+p\to V+p)\int d^2b_{\rm Pb} \frac{xdn_{\rm Pb}(x,b_{\rm Pb})}{dx})S^2_p(b_{\rm Pb})S^2_V(b_{\rm Pb})\,.
\ee 

Calculating the photon flux, $n(x,b)$, given by 
 (\ref{flux0}) we keep just the electric ($F_E$) term
since the magnetic ($F_M$) contribution contains an additional $x^2$ factor, while we work at small $x$. Besides this, the magnetic contribution is concentrated at  low impact parameters where the gap survival factor $S^2_p$ is extremely small. (We have checked that the magnetic ($F_M$) contribution does not exceed 1\%.)
For the lead ion the form factor $F_E$ corresponds to the proton distribution in lead, see~\eqref{rho}.

Note, however, that expression (\ref{flux0}) cannot be transformed to the coordinate, $b$, representation directly. First, we have to transform the {\em amplitude} of photon emission and to account for the polarization structure of the amplitude. The point is that the photon polarization vector $ \vec{e}_\gamma$ is directed parallel to the photon transverse momentum $ \vec{q}_\perp$. That is, the amplitude should be a vector. In $b$ space it will be the vector $ \vec{a}_\gamma= \vec{b}~ a(b)$. Calculating the Fourier transform gives
\be
 \vec{b} ~a(b)=\frac 1{4\pi^2}\int \frac{d^2q_\perp 
e^{i \vec{b} \cdot \vec{q}_\perp}}
{(q^2_\perp+x^2m^2_p)} \vec{q}_\perp\sqrt{\frac{\alpha^{\rm QED}}{\pi x}(1-x)F_E(Q^2)}\,,
\ee 
and after the angular integration we do not obtain the usual zero-order Bessel $J_0(\vec{b}\cdot\vec{q_\perp})$, but rather $J_1(\vec{b}\cdot\vec{q_\perp})$~\cite{Vidovic:1992ik}. Since $J_1(\vec{b}\cdot\vec{q_\perp})$ vanishes as $b\to 0$, the typical values of the impact parameter become larger.

Finally the photon flux in the $b$ representation outside the heavy ion takes the form
\be
\frac{d^3n_{\rm Pb}}{dxd^2b_\gamma} ~=~ \frac{Z^2\alpha^{\rm QED}}{x\pi^2 b_\gamma^2}~ (xm_pb_\gamma)^2~K_1^2(xm_p b_\gamma)\,,
\label{eq:fluxb}
\ee
where $K_1(z)$ is the modified Bessel function.

\subsection{The uncertainty in the evaluation of  $S^2$}
Note that the expression for $S_V$ in (\ref{gap-v}) is written in the spirit of the vector meson dominance (VMD) model~\cite{Sakurai:1960ju, Bauer:1977iq}. That is, we {\em assume} that the photon to $\Upsilon$ transition takes place {\em before} the collision and then the {\em completely dressed} $\Upsilon$ meson interacts with the heavy ion. This is reasonable when the meson goes in the proton beam direction (see~\cite{Khoze:2019xke} for a detailed discussion). However, the assumption is not justified for a very forward (large rapidity) $\Upsilon$ going in the direction of the lead ion. Indeed, the $\gamma\to b\bar b$ vertex is point-like and for $Y>2-3$ the quarks do not have   sufficient time to form the normal $\Upsilon$ wave function.
From the beginning the size of the $b\bar b$ pair is too small, and the corresponding cross section~\cite{Kopeliovich:1981pz, Bertsch:1981py}
\be
\sigma(VN)\propto\alpha^2_s\langle r^2_{b\bar b}\rangle
\ee
is smaller than the cross section of the normally dressed $\Upsilon$ meson.  

In addition, the absorption cross section for the $\Upsilon$ meson, $\sigma(VN)$, is not known experimentally. Moreover it depends on the $VN$ collision energy $s_{VN}=W^2$. For our numerical estimate we take the Regge behaviour
\be
\label{sig-VN}
\sigma_{VN}(W)=\sigma_0\left(\frac{W^2}{M^2_V}\right)^{\alpha_P(0)-1}
\ee
and assume that $\sigma_0\propto 1/M^2_V$. The normalization is fixed to the $J/\psi$ absorptive cross section, $\sigma(J/\psi N)\simeq 4$ mb, at $W\simeq 10$ GeV 
for a completely dressed $J/\psi$ meson measured by the N50 collaboration~\cite{NA50:2006rdp}.
For the Pomeron intercept we take $\alpha_P(0)-1=0.25$. This is consistent with the DIS data and the intercept of the BFKL Pomeron after the resummation of the next-to-leading logarithmic corrections~\cite{Fadin:1998py, Ciafaloni:1998gs, Salam:1998tj, Ciafaloni:1998iv}.
 Since the expected value of $\sigma(\Upsilon N)$ is  small, the survival probability due to $\Upsilon$ absorption, $S^2_V$, is rather close to unity.

To demonstrate the effect, we compare the obtained results with those taking $\sigma(\Upsilon N)=0$. 
The difference never exceeds 6-10\%.
Accounting for the  `undressed meson' problem 
this means that for the mesons going with large rapidity in the lead ion direction the true value of $S^2$ may be about 3-5\% larger.

Finally, there is some uncertainty due to the value of the proton-nucleon cross section $\sigma(pN)$.  Recall that at 8 TeV TOTEM had measured
$\sigma_{\rm tot}(pp)=(103 \pm 2.3)$ mb~\cite{TOTEM:2016lxj},
while ATLAS-ALFA gives $(96 \pm 1)$ mb~\cite{Stenzel:2016kbv}.
Here we take $\sigma(pN)=100$ mb for our numerics. The corresponding uncertainty is about 3-4\%.

\subsection{Numerical results for (photon flux)$\times$(gap survival)}
For proton-lead collisions, eqn.~(\ref{sig-t}) is replaced by the form
\begin{align}
\frac{d\sigma(p+{\rm Pb} \to p+V+{\rm Pb})}{dY}~=
~&S^2(W_{\rm Pb}) \left( k_+ \frac{dn_{\rm Pb} }{dk_+} \right)\sigma_{\gamma p}(W_{\rm Pb}) \nonumber ~\\&+~S^2(W_p) \left( k_- \frac{dn_p}{dk_-} \right) \biggl(\sigma_{\rm incoh}(W_p)+\sigma_{\rm coh}(W_p)\biggr)R^2_A\,,
\end{align}
where the nuclear modification factor $R_A(2\xi)=g_{\rm Pb}(2\xi)/(Ag_p(2\xi))$ accounts for the fact that the gluon distribution\footnote{Recall that the value of $\sigma(\gamma+p\to V+p)\propto (2\xi\, g(2\xi))^2$ is almost completely driven by the gluon distribution.} in the lead ion may differ from the sum of the gluon distributions in free nucleons. The value of the nucleon modification factor $R_A(2\xi)$ at the corresponding scale $\mu=
M_V/2$ is taken from the EPPS16 NLO analysis~\cite{Eskola:2016oht}.
It is convenient to introduce the so-called `effective fluxes',
$f_{\rm Pb}$ and $f_p$, which include the original photon flux $xdn/dx$ times the survival and nuclear modification effects\footnote{It is not completely correct to use the word `flux' for a quantity which includes these additional effects; however we use it since it enables us to shorten the description of the computations.}. The effective flux radiated by the lead ion is
\be
\label{f-Pb}
f_{\rm Pb}(2\xi)=\frac{\sigma_{\rm Pb}}{\sigma(\gamma+p\to V+p)}\,,
\ee
where $\sigma_{\rm Pb}$ is given by (\ref{sig-Pb}).  The effect of the flux radiated by the proton is given by the sum of coherent and incoherent fluxes, 
\be
\label{f-incoh}
f_{\rm incoh}(2\xi)=\frac{\sigma_{\rm incoh}}{\sigma(\gamma+p\to V+p)}R^2_A(2\xi)\ 
\ee
and
\be
\label{f-coh}
f_{\rm coh}(2\xi)=\frac{\sigma_{\rm coh}}{\sigma(\gamma+p\to V+p)}R^2_A(2\xi)\,,
\ee
where $\sigma_{\rm incoh}$ and  
 $\sigma_{\rm coh}$ are given by (\ref{sig-inc}) and (\ref{sig-coh}), respectively.  Here, $2\xi = 2 M_V^2/(2W^2 - M_V^2) \approx M_V^2/W^2$ for $W^2 \gg M^2_V$, i.e. $2\xi \ll 1$, is the fractional momentum transfer provided by the two-gluon exchange.
 
Finally, the cross section of heavy vector meson production can be written as 
 \be
 \label{sig-f}
 \frac{d\sigma(p+{\rm Pb} \to p+V +{\rm Pb})}{dY}~
=f_{\rm Pb}(W_{\rm Pb})\sigma_{\gamma p}(W_{\rm Pb})~+~\biggl(f_{\rm incoh}(W_p)+f_{\rm coh}(W_p)\biggr)\sigma_{\gamma p}(W_p)\,.
 \ee 
Here the values of $W_{\rm Pb}$ and $W_p$ correspond to the energies of the $\gamma+p\to V+p$ process initiated by the photon emitted off the lead ion and proton beam, respectively.

The values of the effective fluxes are presented in Tables 1 and 2 in Appendix B for $\Upsilon$ and $J/\psi$ production at proton-nucleon collision energy $\sqrt{s_{pN}}=5.02$ TeV. The analogous values at $\sqrt{s_{pN}}=8.16$ TeV are given in Tables 3 and 4.
 
\section{Comparison with data}
In this section, we compare our theoretical predictions with the rapidity differential cross section data from ALICE~\cite{ALICE:2014eof,
ALICE:2018oyo} and CMS~\cite{CMS:2018bbk} for the process $p+{\rm Pb} \to p+V +{\rm Pb}$, where $V = J/\psi, \Upsilon$.  Using eqn.~\eqref{sig-f} we compute $\text{d} \sigma (p+{\rm Pb} \to p+J/\psi +{\rm Pb})/\text{d}Y$, with $\sigma_{\gamma p}$ evaluated using, as input, the gluon PDF fit obtained from our previous analyses~\cite{Flett:2020duk}. 

\begin{figure} [h]
\centering
\includegraphics[width=0.4\textwidth]{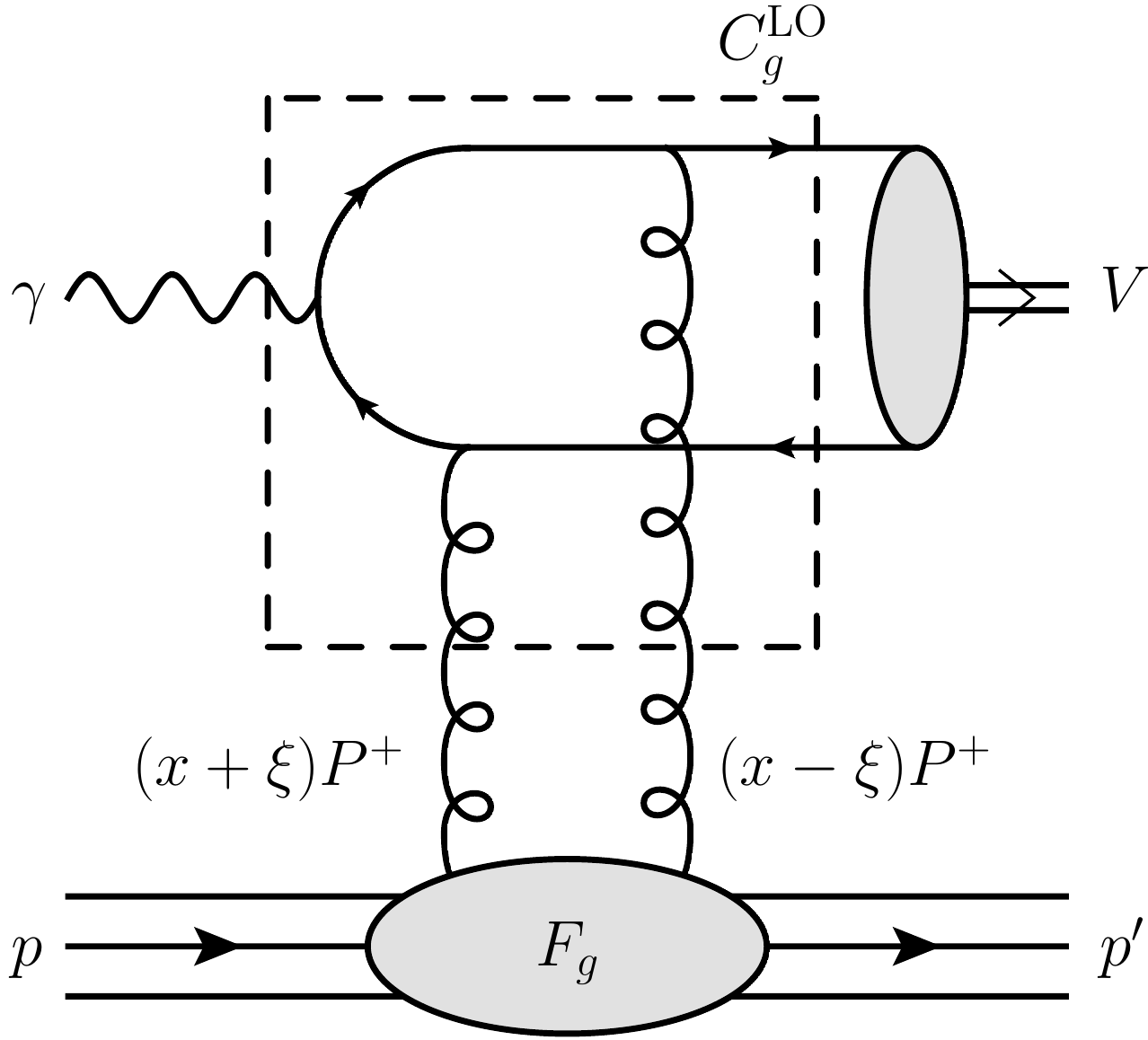}
\qquad
\includegraphics[width=0.4\textwidth]{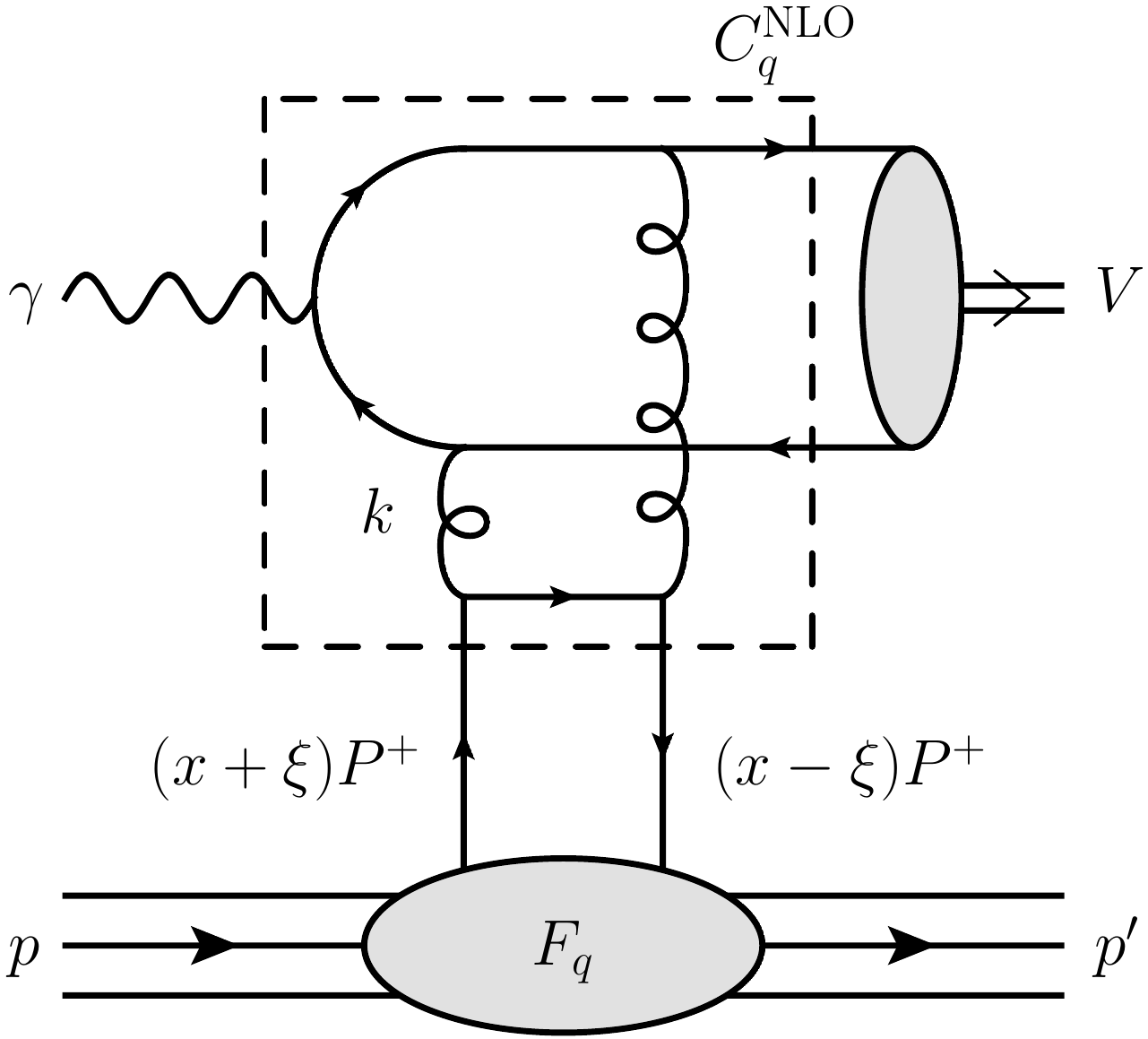}
\caption{\sf{
 The LO and the NLO quark contributions to $\gamma+p\to V+p$ amplitude. The parton momentum fractions are $x+\xi$ and $x-\xi$ and $k$ is the loop momentum.}}
\label{f1-l4}
\end{figure}

We work at NLO within 
the collinear factorisation scheme and express the amplitude for exclusive heavy vector meson photoproduction as
\begin{equation}
   \mathcal A = \frac{4 \pi \sqrt{4 \pi \alpha} e_q ( \epsilon_V^* \cdot \epsilon_\gamma)}{N_c} \left(\frac{8\,\langle O_1 \rangle_V}{M_{V}^3} \right)^{1/2} \int_{-1}^1 \text{d}x \biggl( C_g(x,\xi) F_g(x,\xi) + C_q(x,\xi) F_q(x,\xi) \biggr)\,,
    \label{amp}
\end{equation}
where $F_g$ and $F_q$ are the gluon and quark singlet Generalised Parton Distributions~(GPDs), $C_g$ and $C_q$ are the gluon and quark coefficient functions, see~\cite{Ivanov:2004vd, Jones:2016ldq}, and $x-\xi,~x+\xi$ are the parton momentum fractions in the lightcone direction~$P^+$. The corresponding gluon and quark coefficient functions for exclusive heavy vector meson electroproduction were calculated in~\cite{Chen:2019uit, Flett:2021ghh}. The dependence on the factorisation and renormalisation scales, $\mu_F, \mu_R$, and on the four-momentum transfer squared, $t$, is not shown. The set-up is illustrated in Fig.~\ref{f1-l4}. The non-relativistic QCD~(NRQCD) matrix element $\langle O_1 \rangle_{V} $ is fixed by the experimental value of the heavy vector meson decay width to a dilepton pair, see~\cite{Bodwin:1994jh}. 

We take $\mu_R=\mu_F$ and use the `optimal' factorization scale $\mu_F=M_ V/2$, see~\cite{Jones:2015nna}. That is, the exclusive $J/\psi$ and $\Upsilon$ photoproduction data probe the gluon densities at two different scales. Besides this, we implement the $Q_0$ subtraction~\cite{Jones:2016ldq} needed to avoid the double counting between the low $k_T<Q_0$ contributions in the NLO coefficient functions and that hidden in the PDF inputs. This choice provides a sufficiently good scale stability of the prediction\footnote{Recall that exactly the same approach (that is use of the optimal scale $\mu_F=M_V/2$ and the $Q_0$ subtraction) was applied in~\cite{Flett:2020duk} to determine the low-$x$ gluon PDF from exclusive $J/\psi$ data, used in the present analysis.}.

The cross section, $\sigma_{\gamma p}$, integrated over the Mandelstam variable $t$, is given by 
\begin{equation}
\sigma_{\gamma p}(W) = \frac{1}{B_V(W)} \left( \frac{\text{d} \sigma}{\text{d} t} \left(\gamma p \rightarrow V p\right)\biggl|_{t=0} \right) =  \frac{1}{B_V(W)} \frac{(\text{Im} \mathcal A)^2 (1 + \rho^2)}{16 \pi W^4 }\,,
\end{equation}
where 
\be
\rho ~~=~~ \frac{{\rm Re}\mathcal A}{ {\rm Im} \mathcal A}~~=~~{\rm tan}\left(\frac{\pi}{2}~\frac{\partial \ln ({\rm Im} \mathcal A/W^2)}{\partial \ln W^2}\right)
\ee
is the real part correction, see e.g.~\cite{Ryskin:1995hz}, and $B_V(W)$ is the Regge-motivated energy dependent slope parameter given by
\begin{equation}
B_V(W) = \left (B_0 + 4 \alpha'_P \ln \left( \frac{W}{W_0} \right) \right)\,\,\,\text{GeV}^{-2}\,.
\end{equation}
Here, $B_0 = 4.9$ for $V = J/\psi$~\cite{H1:2002yab} and $B_0 = 4.63$ for $V = \Upsilon$, with $\alpha'_P = 0.06$ and $W_0 = 90$~GeV~\cite{Khoze:2013dha}. The smaller value of $B_0$ for the case $V = \Upsilon$ is due to the need for reducing the $J/\psi$ slope parameter by the factor $4 \alpha' \ln \left(M_{\Upsilon}/M_{J/\psi}\right)$.

    \begin{figure*}
        \centering
        \begin{subfigure}[b]{0.496\textwidth}
            \centering
            \includegraphics[width=\textwidth]{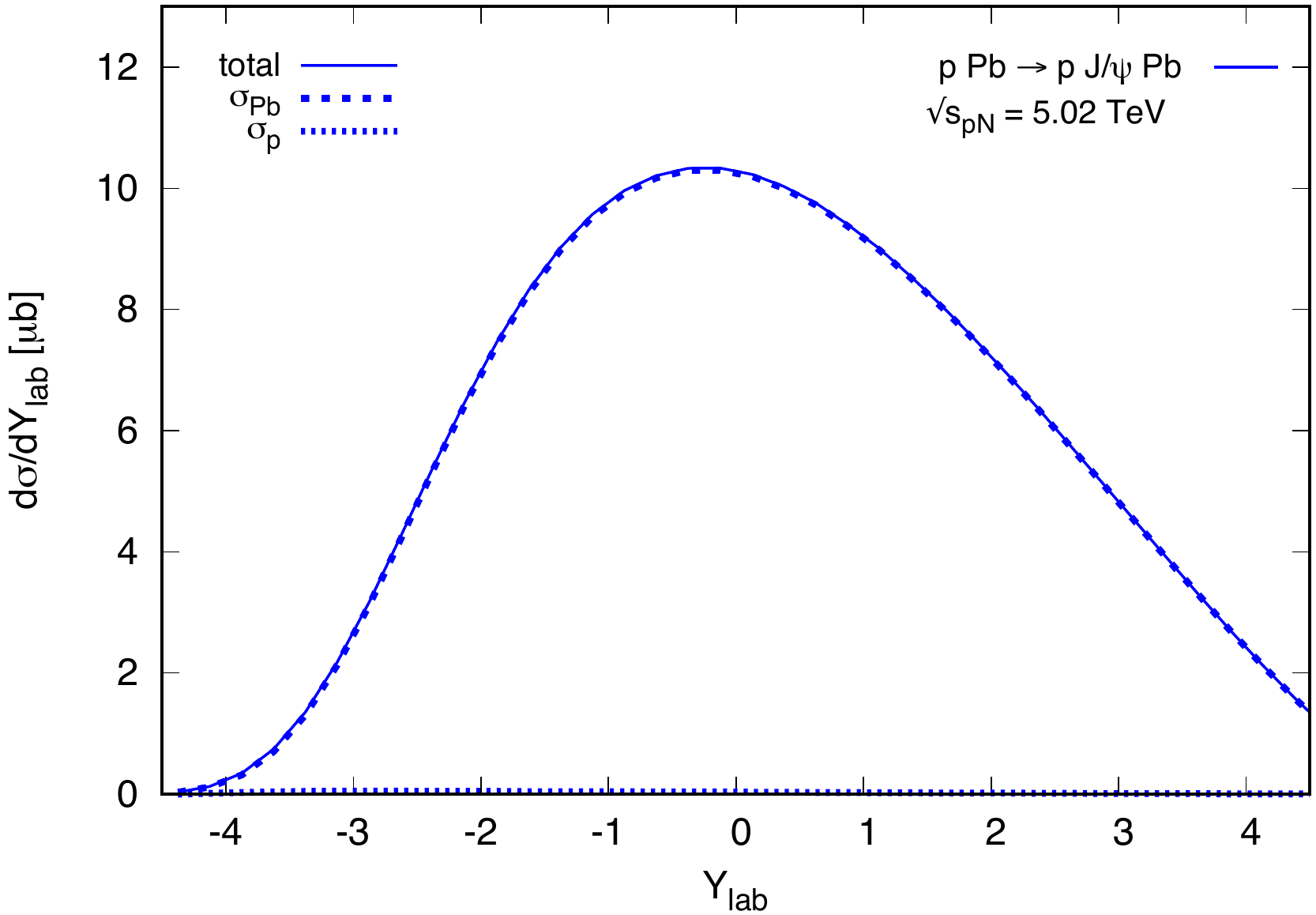}
            \caption[]%
            {{\small}}    
            \label{fig:jpsi5decomps}
        \end{subfigure}
        \hfill
        \begin{subfigure}[b]{0.496\textwidth}  
            \centering 
            \includegraphics[width=\textwidth]{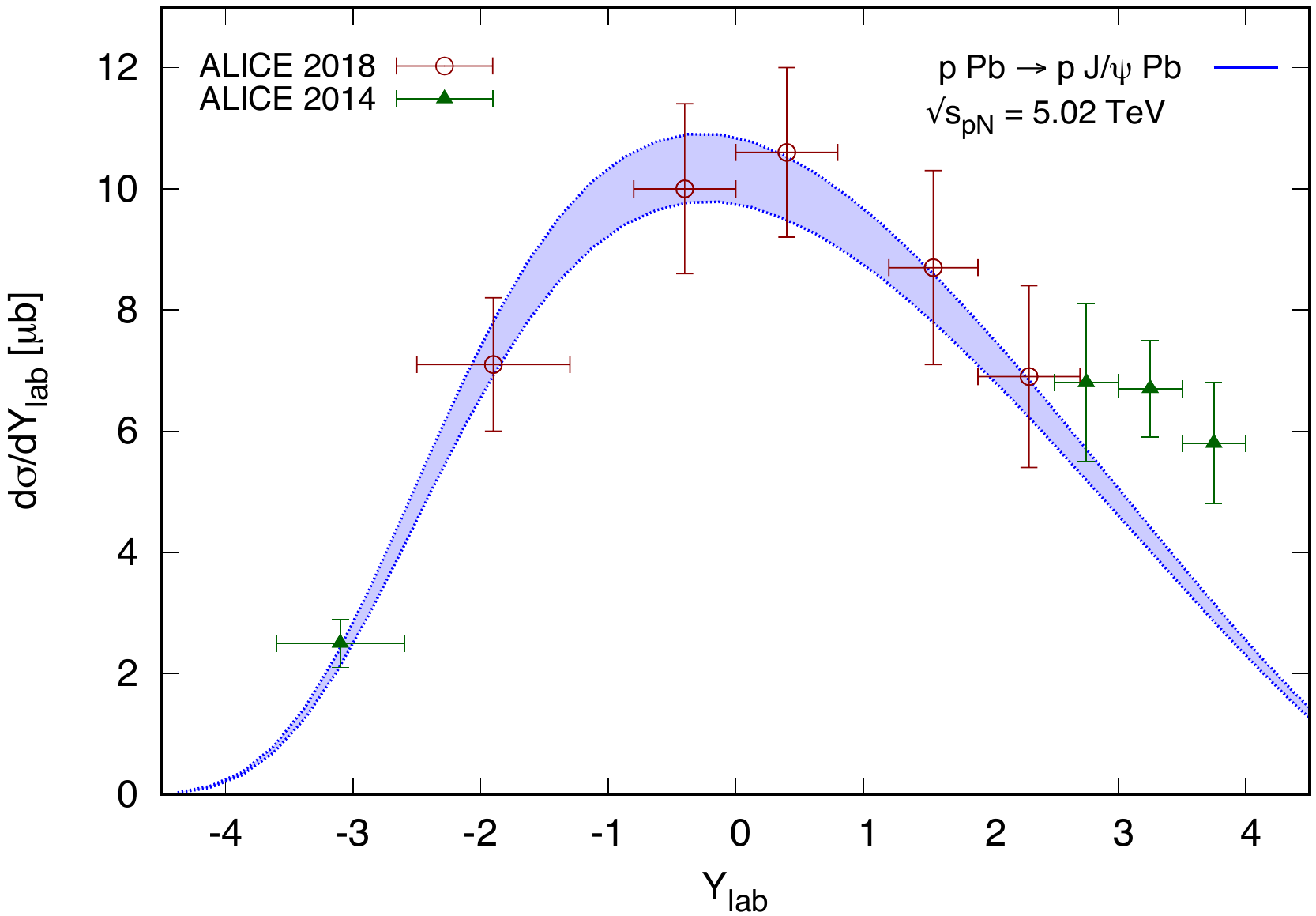}
            \caption[]%
            {{\small }}    
            \label{fig:jpsi5data}
        \end{subfigure}
        \vskip\baselineskip
        \begin{subfigure}[b]{0.496\textwidth}   
            \centering 
            \includegraphics[width=\textwidth]{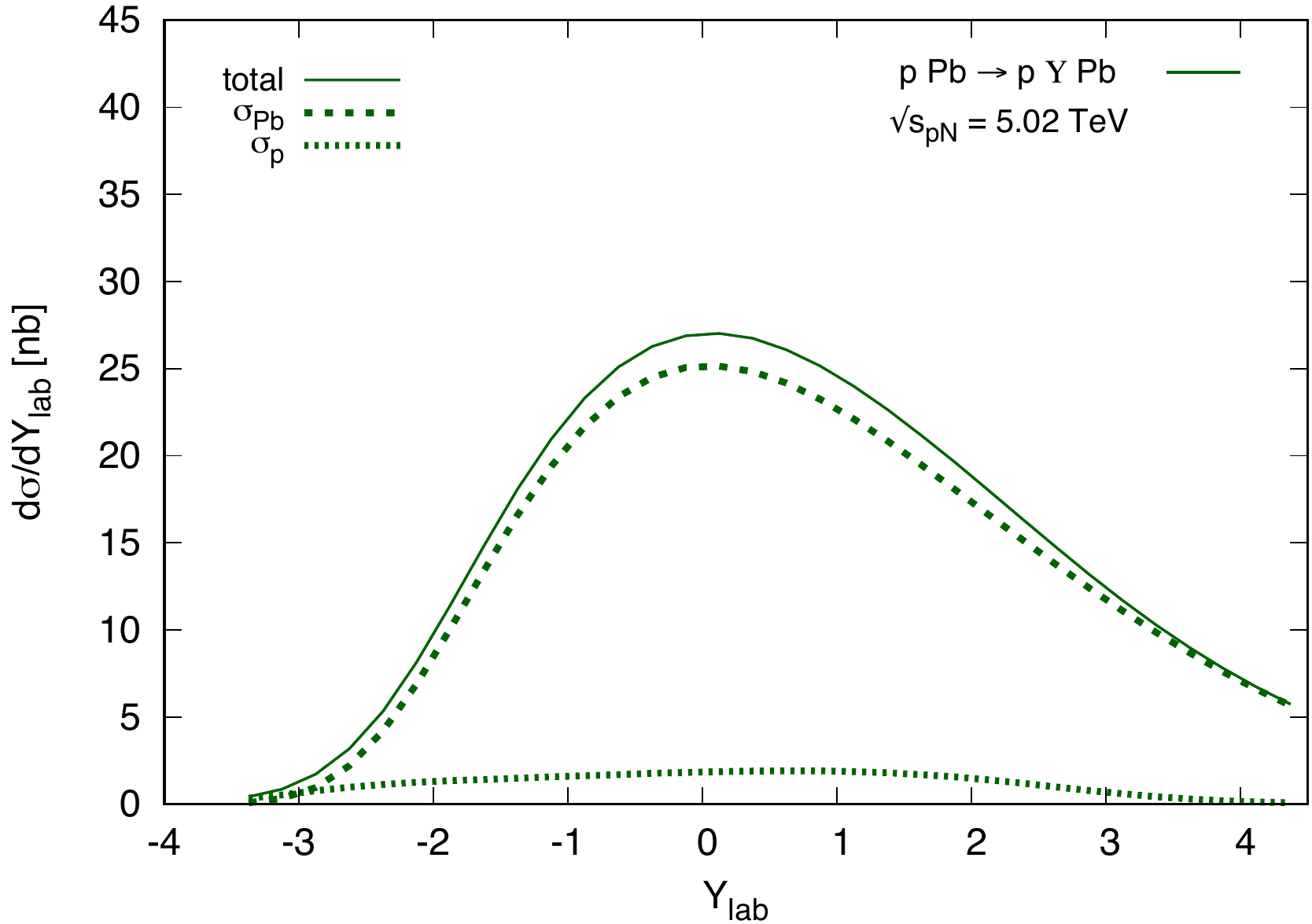}
            \caption[]%
            {{\small }}    
            \label{fig:ups5decomps}
        \end{subfigure}
        \hfill
        \begin{subfigure}[b]{0.496\textwidth}   
            \centering 
            \includegraphics[width=\textwidth]{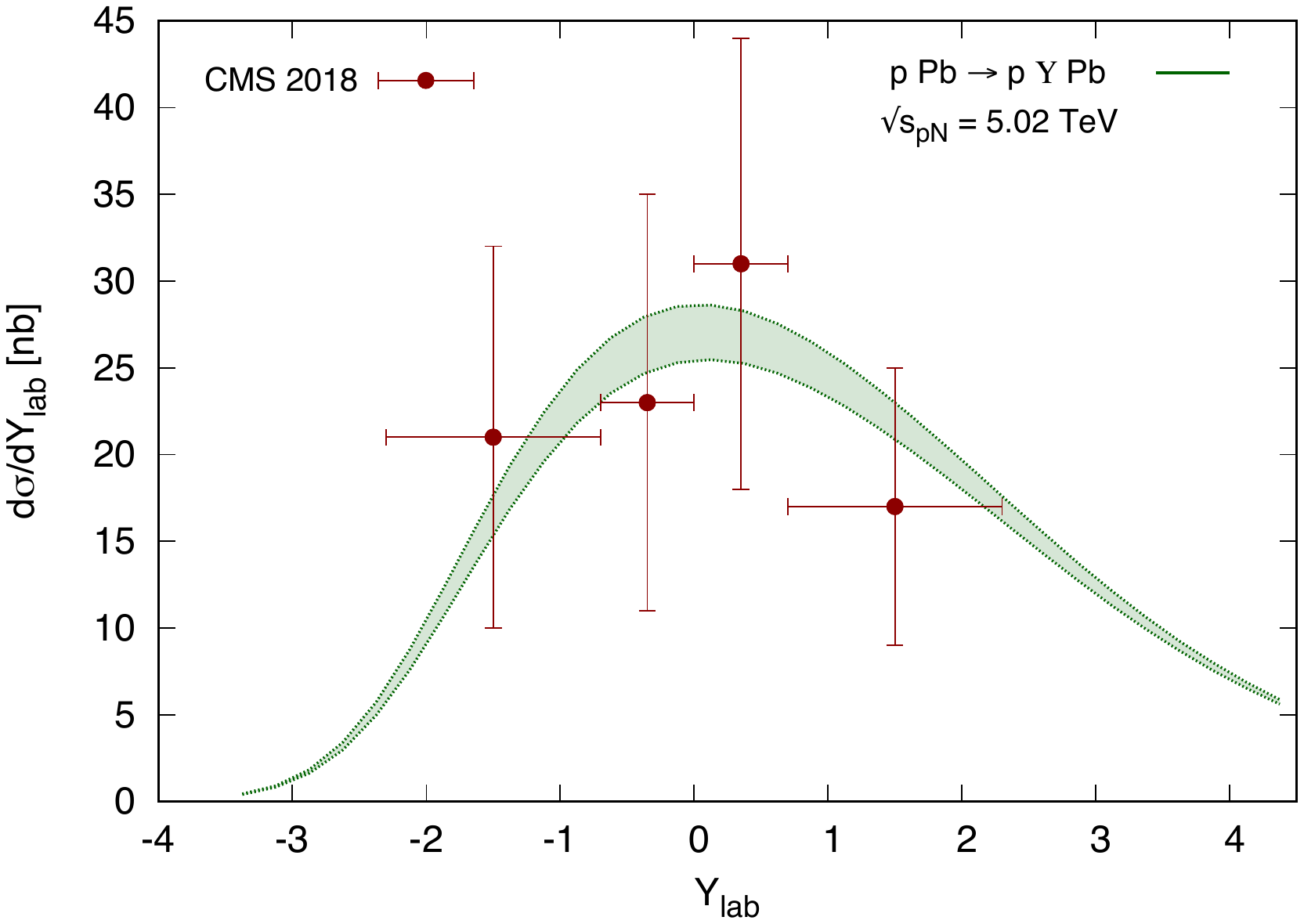}
            \caption[]%
            {{\small }}    
            \label{fig:ups5data}
        \end{subfigure}
        \caption[  ]
        {\sf{Theoretical predictions for coherent exclusive $J/\psi$ and $\Upsilon$ photoproduction rapidity differential cross sections in $p$Pb collisions with $\sqrt{s_{pN}} = 5.02$ TeV. Panels (a) and (c) show the decomposition of the total cross section into the $\sigma_{\rm Pb}$ and $\sigma_p$ contributions, see text for details. Panels (b) and (d) compare our predictions with existing data from ALICE and/or CMS~\cite{ALICE:2014eof, ALICE:2018oyo, CMS:2018bbk}. 
        We emphasise that no fit is made to the data shown, the width of the uncertainty bands is obtained by the $\pm 1 \sigma$ statistical uncertainties and the normalisation errors of the datasets used in our previous fit~\cite{Flett:2020duk} to determine the low $x$ gluon PDF, which is used in the present analysis.}} 
        \label{fig:5tev}
    \end{figure*}

Fig.~\ref{fig:5tev} shows our rapidity differential cross section predictions at $\sqrt{s_{pN}} = 5.02$ TeV for the process $p+{\rm Pb} \to p+V +{\rm Pb}$. The upper row shows the predictions for $V = J/\psi$ and the lower one for $V = \Upsilon$. In the left panels, we show the decomposition of the total cross section result into the $\gamma p$ contribution (the first term on the right hand side of eqn.~\eqref{sig-f}, labelled by $\sigma_{\text{Pb}}$ in the figure) and the $\gamma$Pb contribution (the second term on the right hand side of eqn.~\eqref{sig-f}, labelled by $\sigma_{p}$ in the figure). In the right panels, we show our results compared with existing data from ALICE and/or CMS~\cite{ALICE:2014eof, ALICE:2018oyo, CMS:2018bbk}. 
The error bands are indicative of the uncertainty due to our previous low $x$ gluon PDF fit~\cite{Flett:2020duk} only and does not account for theoretical uncertainties in the present formalism. We emphasise that no fit is performed to the $p$Pb data here, and the width of the bands is given by propagating the errors on the fit parameters using the full covariance matrix obtained from our previous gluon PDF fit made to the low $x$ exclusive $J/\psi$ data in $pp$ collisions in~\cite{Flett:2020duk}. 

For $V=J/\psi$, the $\gamma$ Pb contribution is less than a percent at mid-rapidity, while for $V = \Upsilon$, it is as much as 7\%. The mass
of the $\Upsilon$ is $\sim 3$ times that of the $J/\psi$ and so (with $k_+ \propto M_V$ and $W_+ \propto \sqrt{M_V}$), the typical photon energy in exclusive $\Upsilon$ production is now much larger than in exclusive $J/\psi$ production for a given $Y_{\text{lab}}$. Therefore, the photon flux radiated off Pb is relatively suppressed for $\Upsilon$ production.  We remark that for $Y_{\text{lab}} < -3$ in Fig.~\ref{fig:ups5decomps}, the photon radiated by Pb has a somewhat greater $x$ (with respect to the nucleon in Pb) and so the value of $b_{\text{Pb}}=b_\gamma$ (in eq.~\eqref{eq:fluxb}) is rather small. This means that the integrated flux from Pb is not large and $S^2(b_{\text{Pb}})$ is small.
On the other hand, the photon radiated by the proton has a much smaller $x$ (with respect to the proton) leading to a larger $S^2$ and a larger integrated flux. This is evident from Tab.~\ref{Tab_1} where, for $Y_{\text{lab}} < -3$, the value of $f_{\text{Pb}}$ is an order of magnitude smaller than $f_p$. The energy dependence of $\sigma(\gamma+N\to V+N)$ does not compensate this difference for $\sqrt{s_{pN}} = 5.02$~TeV. This accounts for the $W_p$ contribution being greater than the $W_{\text{Pb}}$ one in this region.

Our predictions agree favourably with the ALICE data, particularly at backward, central and semi-forward rapidities. For $V=J/\psi$, our prediction undershoots the data for $Y_{\text{lab}} \geq 2.5$. In this region, however, our prediction is well constrained. As shown above, the dominant contribution comes from the photon radiated by Pb, occuring at large $b$, with $S^2$ very close to unity where we have practically no uncertainty in the photon flux. On the other hand, the `gamma-proton' centre of mass energy is relatively small, and this is precisely the
region where our gluon PDF fit agreed well with the HERA data~\cite{Flett:2020duk}. The earlier ALICE data from~\cite{ALICE:2014eof} for $Y_{\text{lab}} \geq 2.5$ show an unexpected plateauing behaviour and are arguably inconsistent with those in~\cite{ALICE:2018oyo}\footnote{ 
The behaviour of the earlier ALICE data in Fig. 6b are incompatible with other theoretical predictions, too, see e.g. Fig. 4 of~\cite{Guzey:2013taa}.}.

\begin{figure} [h!]
\begin{center}
\includegraphics[width=0.61\textwidth]{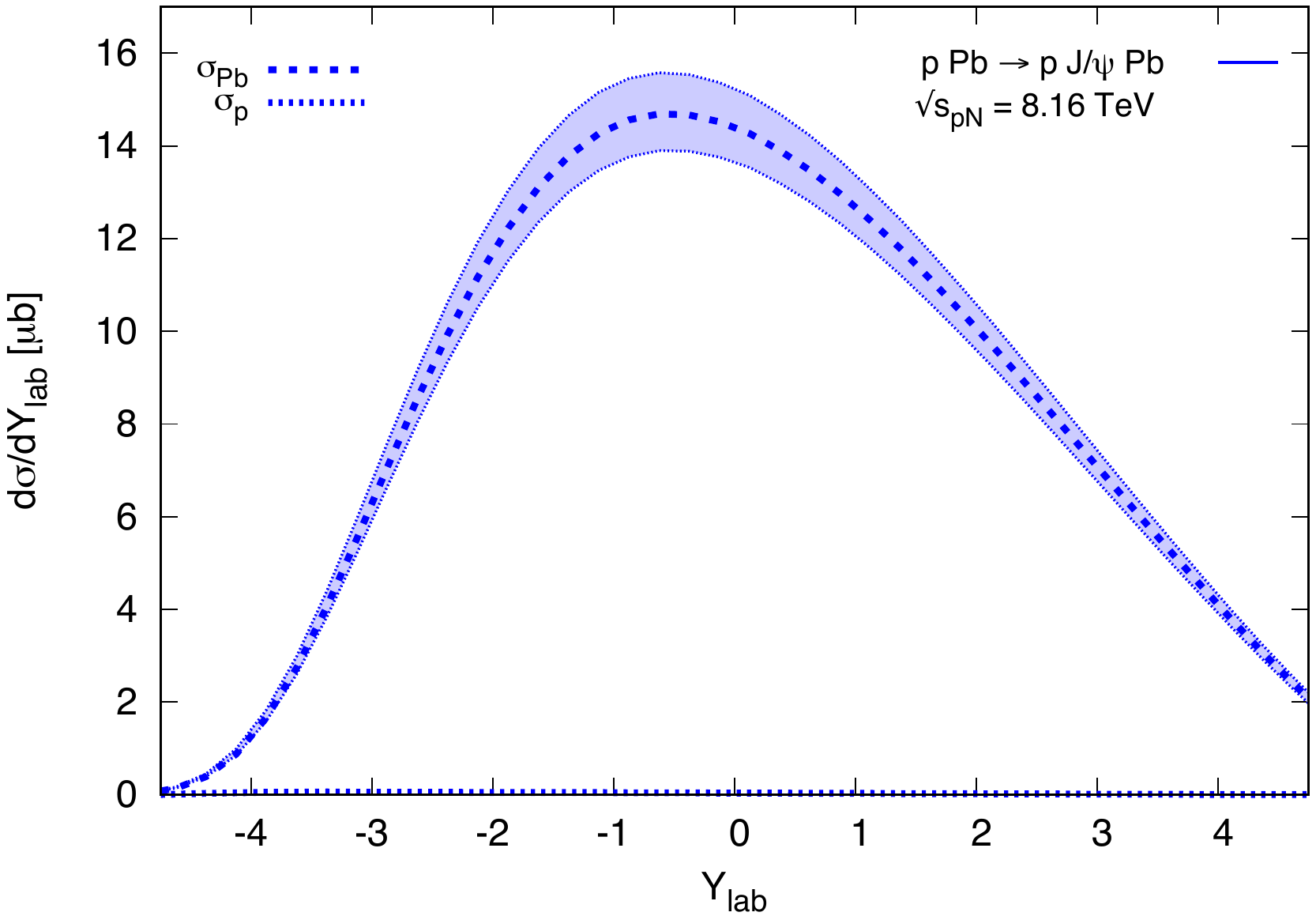}
\vfill
\vspace{0.5cm}
\includegraphics[width=0.61\textwidth]{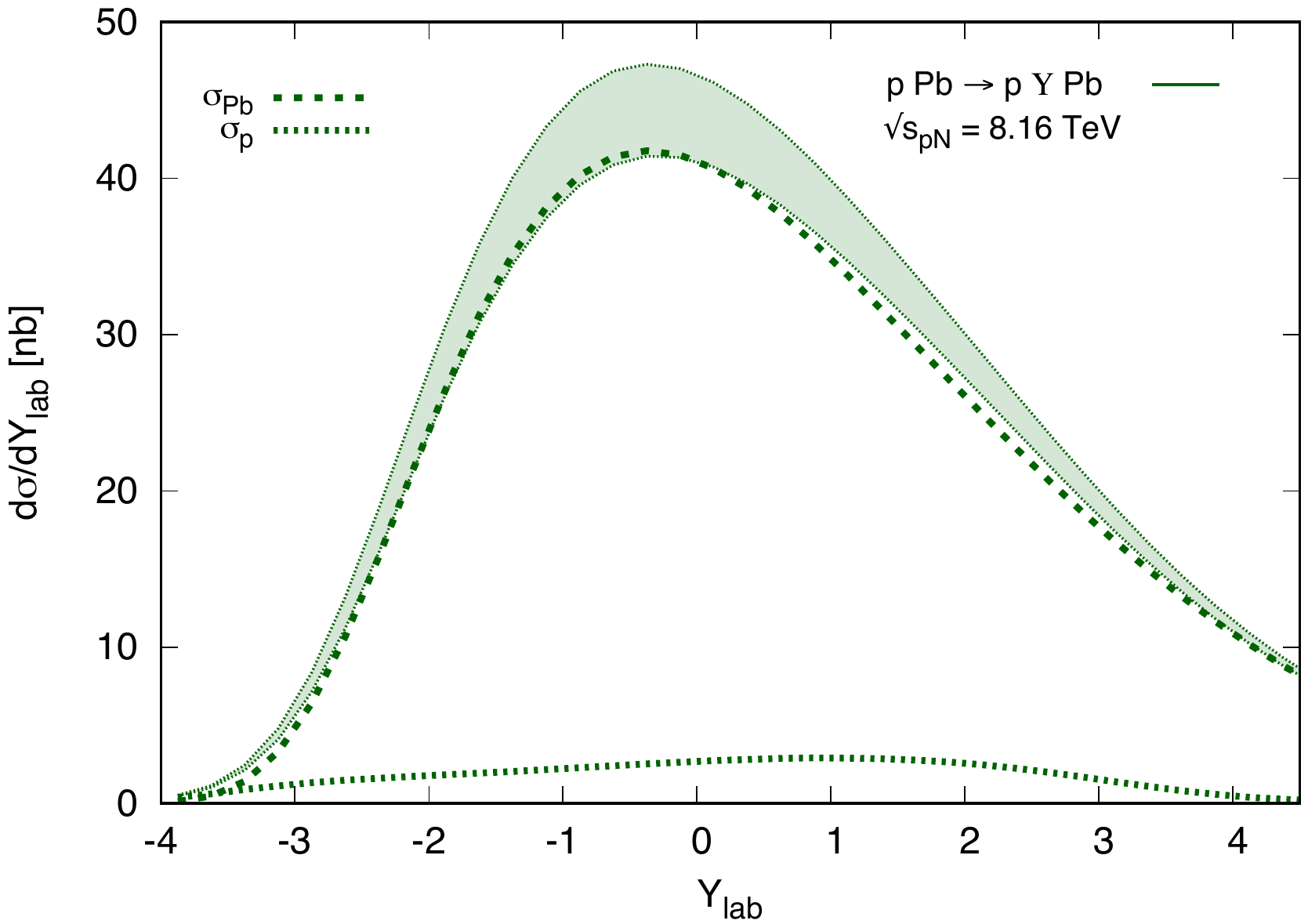}
\caption{\sf{Theoretical predictions for coherent exclusive $J/\psi$ and $\Upsilon$ photoproduction rapidity differential cross sections in $p$Pb collisions with $\sqrt{s_{pN}} = 8.16$ TeV. The bands and curves are as described in the caption of Fig.~\ref{fig:5tev}. Note that currently there are no data for the process $p+{\rm Pb} \to p+V +{\rm Pb}$, with $V = J/\psi, \Upsilon$, at this centre-of-mass energy, but measurements for $\Upsilon$ photoproduction are shortly anticipated~\cite{Dutta:2017izs, Naskar:2018frq}.}}
\label{fig:8tev}
\end{center}
\end{figure}

In Fig.~\ref{fig:8tev}, we show the analogous set of plots at $\sqrt{s_{pN}} = 8.16$ TeV. There are currently no data for $p+{\rm Pb} \to p+V +{\rm Pb}$ at this centre-of-mass energy, but forthcoming measurements for the case $V=\Upsilon$ from CMS are anticipated~\cite{Dutta:2017izs, Naskar:2018frq}. Our results for the rapidity distribution are qualitatively consistent with the shapes predicted by other approaches, however, the normalisation of our prediction is larger than that obtained from CGC models~\cite{Iancu:2003ge, Kowalski:2006hc}. 

In both Figs.~\ref{fig:5tev} and~\ref{fig:8tev}, note that the larger $\text{d}\sigma/\text{d}Y_{\text{lab}}$ at positive $Y_{\text{lab}}$ is indicative of the proton direction corresponding to the positive $Y_{\text{lab}}$. At negative $Y_{\text{lab}}$ the distribution is skewed due to a smaller energy per nucleon in the Pb-ion. Indeed, at large rapidities in the Pb direction (i.e. large negative $Y_{\text{lab}}$), the contribution due to the photon radiated from Pb is strongly suppressed, especially in the $\Upsilon$ case, since the momentum fraction, $x\propto (M_V/\sqrt{s})\exp(Y)$ carried by the photon becomes large and the transverse momentum cutoff $xm_p$ becomes  comparable with the inverse lead ion radius $1/R_{\text{Pb}}$, i.e. here we deal with the impact parameter $b_{\text{Pb}}\sim R_{\text{Pb}}$. Therefore, in this region, the survival factor $S^2(b_{\text{Pb}})$ suppresses the $\sigma_{\text{Pb}}$ contribution, and the second term of~\eqref{sig-f}, i.e. the photons radiated by the proton, starts to dominate.

\section{Conclusions}

We have predicted the cross sections for exclusive $J/\psi$ and $\Upsilon$
meson production in proton-lead ion collisions at the LHC for centre-of-mass energies $\sqrt{s_{pN}} = 5.02$ TeV and $\sqrt{s_{pN}} = 8.16$ TeV, using the low $x$ gluon distribution extracted in~\cite{Flett:2020duk} from
the data on $p+p\to p+J/\psi+p$.
We account for the gap survival probability caused by both the additional
proton-lead interactions and the interaction of the secondary vector
meson inside the heavy ion.

As expected, the dominant contribution to the cross section in exclusive $p$Pb collisions comes from the amplitude where the photon is radiated by the lead ion (for all rapidities $Y_{\text{lab}} \gapproxeq{-3}$).  However, our detailed study finds that the naively expected dominance of the $Z^2$ enhancement of the cross section, 
when we go from exclusive production from photons radiated by a proton to that for photons radiated by the heavy ion, is considerably reduced. 
Nevertheless, we find that the enhancement is still sufficient to enable forthcoming data for exclusive heavy vector meson production in $p$Pb collisions to provide additional constraints
on the {\it free} proton gluon PDF at low to moderate values of $x$ and scale. Such data could be used in a combined fit with those in $pp$ collisions anticipated from the High-Luminosity phase of the LHC, as well as in the upcoming $ep$ programme of the Electron-Ion collider, to provide refined constraints on the gluon PDF. 

\vspace{1cm}
{\bf\large\noindent Appendix A: 
Optical density of the Pb ion}

The optical density of the Pb ion may be written in the form
\begin{equation}
\label{T-opt}
T(b_{\rm Pb})=\int_{-\infty}^{+\infty}dr_z\,(\rho_p(r)+\rho_n(r)),
\end{equation}
with $r=\sqrt{r_z^2+r^2_t}$. For the nucleon density in lead, $\rho(r)$, we use 
the Woods-Saxon form~\cite{Woods:1954zz}
\be
\label{rho}
\rho_N(r)= \frac{\rho_0}{1+\exp{((r-R)/d)}}\;,
\ee
where the parameters  $d$ and $R$, respectively, characterise the skin thickness and the radius of the nucleon density in the heavy ion; $r=(r_z,r_t)$. For $^{208}$Pb
we take the recent results of~\cite{Tarbert:2013jze,Jones:2014aoa}
\begin{align}\nonumber
R_p &= 6.680\, {\rm fm}\;, &d_p &= 0.447 \, {\rm fm}\;,\\ \label{eq:pbpar}
R_n &= (6.67\pm 0.03)\, {\rm fm}\;, &d_n &= (0.55 \pm 0.01) \, {\rm fm}\;.
\end{align}
The nucleon densities, $\rho$, are normalized to 
\be
 \int\rho_p(r)d^3r=Z \;, \qquad \int\rho_n(r)d^3r=N_n\;,
\ee
for which the corresponding proton (neutron) densities are $\rho_0 = 0.063$ (0.093) ${\rm fm}^{-3}$.

\vspace{1cm}

{\bf\large\noindent Appendix B: 
Tables of effective fluxes}

The values of the effective fluxes used in eqn.~\eqref{sig-f} are presented in Tables 1 and 2 below for $\Upsilon$ and $J/\psi$ production at proton-nucleon collision energy $\sqrt{s_{pN}}=5.02$ TeV. The analogous values at $\sqrt{s_{pN}}=8.16$ TeV are also given, in Tables 3 and 4.

The first column is the rapidity, $Y_{\rm lab}$, of the vector meson measured in the laboratory frame. We account for the asymmetry of proton-ion collisions. In the case of lead we have (in the laboratory frame) a proton beam momentum equal to 4 TeV (6.5 TeV), 
while the momentum of a nucleon in lead is $4~(Z/A) = 1.58~$TeV ($6.5~(Z/A) = 2.56~$TeV) for $\sqrt{s_{pN}} = 5.02$ (8.16) TeV. Positive $Y_{\rm lab}>0$ corresponds to the vector meson going in the proton direction.

The second and fourth columns are the $\gamma$-nucleon collision energies $W_{\rm Pb}$ and $W_p$.
The effective flux $f_{\rm Pb}$ is given in the third column while the fluxes $f_{\rm incoh}$, $f_{\rm coh}$ and their sum
are shown in the fifth, sixth and seventh column, respectively. 

\begin{figure} [h!]
\centering
\includegraphics[width=0.47\textwidth]{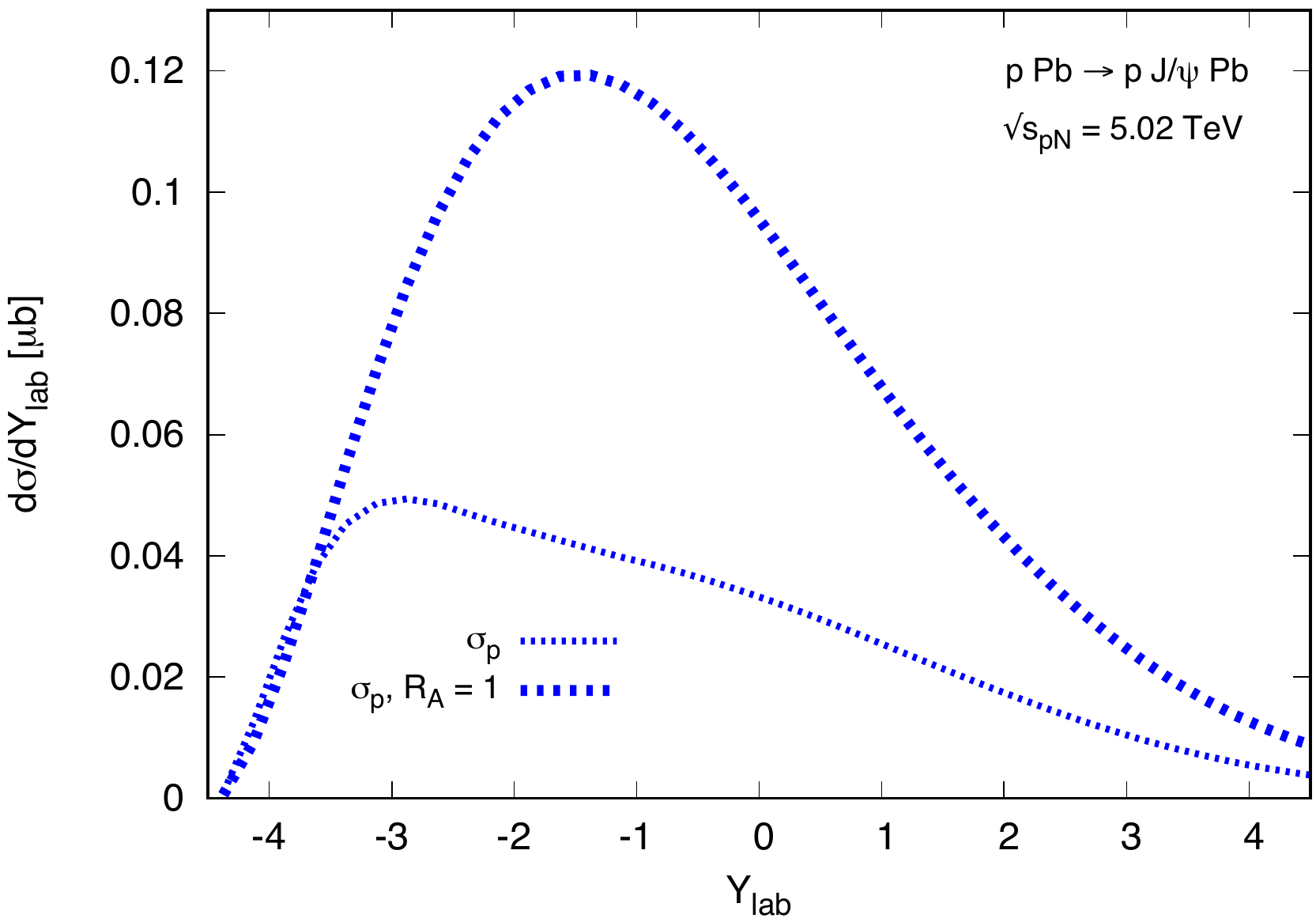}
\qquad
\includegraphics[width=0.47\textwidth]{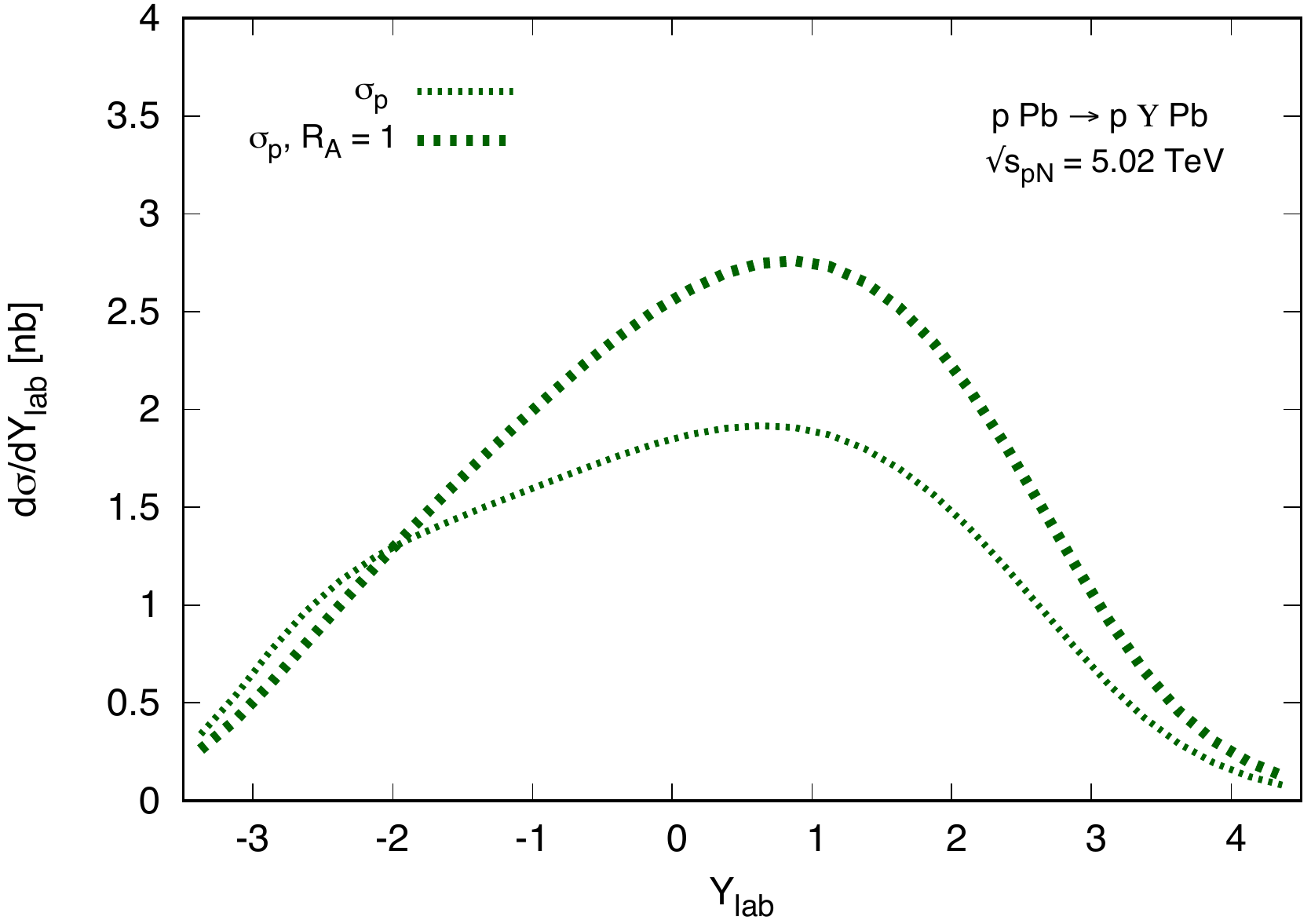}
\caption{\sf{The contribution due to the term where the photon is emitted from the proton, $\sigma_p$, with and without inclusion of the nuclear modification factor, $R_A$. The left panel is for the $J/\psi$ production at $\sqrt{s_{pN}} = 5.02$ TeV and the right panel is for $\Upsilon$ production at the same centre-of-mass energy.}}
\label{fig:RA}
\end{figure}

Column eight gives the sum $f_{\rm incoh} + f_{\rm coh}$ for the case $R_A = 1$. The factor of $R_A$ enters only the term where the photon is radiated by the proton and, as shown in Figs.~\ref{fig:5tev} and~\ref{fig:8tev}, this contribution is
relatively small (especially for $J/\psi$) and can be seen at large backward rapidities only. The contribution due to the term where the photon is radiated from the proton with and without inclusion of the nuclear modification factor, $R_A$, is shown in Fig.~\ref{fig:RA}. This difference at the total cross section level amounts to at most a few percent only for the $J/\psi$ case because the contribution when the photon is radiated from the lead ion dominates. Therefore, the nuclear modification factor is hardly seen in $p$Pb collisions, where the first term of~\eqref{sig-f} dominates. This factor can be
much better observed (and measured) in the Pb-Pb set-up, see e.g.~\cite{Eskola:2022vpi} for a baseline description of the Pb + Pb $\rightarrow$ Pb + $J/\psi$ + Pb process in the collinear factorisation framework to NLO.

\begin{figure} [t]
\centering
\includegraphics[scale=0.7]{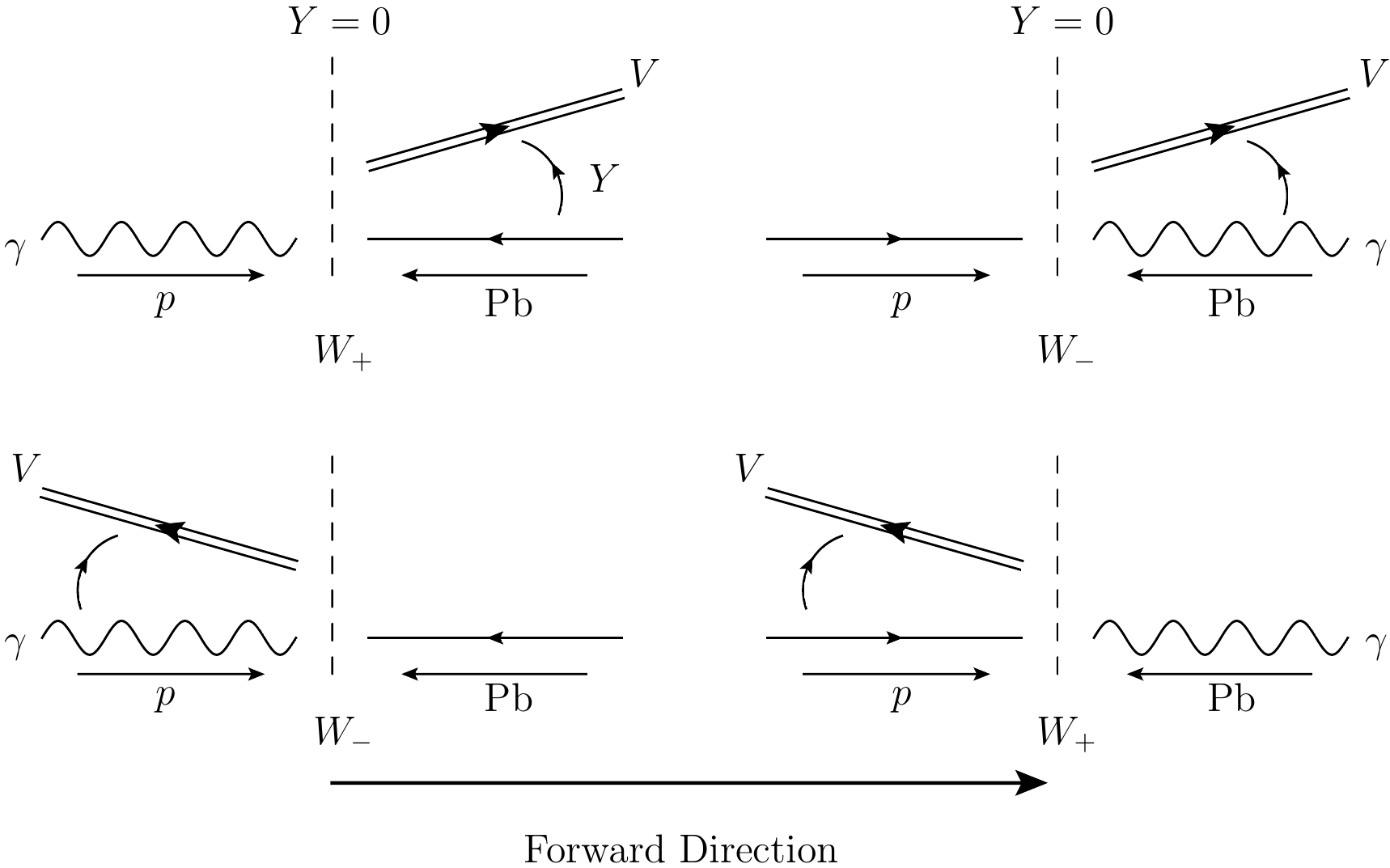}
\begin{center}
\caption{\sf{The four kinematic configurations for exclusive heavy vector meson, $V$, ultraperipheral production in $p$Pb collisions at a given rapidity $Y$ occurring via a photoproduction subprocess at two energies, $W_+$ and $W_-$. The incoming photon is a Weizs\"{a}cker-Williams photon that has been emitted from a proton (lead nucleus) and interacts with an incoming lead nucleus (proton). The vector meson can travel in the direction of the photon ($W_+$), or against the direction of the photon ($W_-)$.}}
\label{configs}
\end{center}
\end{figure}

Note that in asymmetric $p$Pb collisions, the laboratory frame does not coincide with the centre of momentum frame of the $p$Pb system and so eqn. (2) should be adjusted. For the case where the vector meson is detected at positive rapidities (corresponding to the configurations in the upper row of Fig.~\ref{configs}),

\begin{equation}
\label{W}
W^2_+=M_V\sqrt s~e^{+ (Y_{ \text{lab}}-Y_0)} = W_p^2\;\;\;\mbox{and}\;\;\; W^2_-=M_V\sqrt s~e^{- (Y_{ \text{lab}}-Y_0)} = W_{\text{Pb}}^2\,,
\end{equation}
where $Y_0 = 0.465$ is the shift of the nucleon-nucleon centre of mass with respect to the laboratory frame.

\begin{table}[]
\centering
\vspace{-2cm}
\scalebox{.9}{
\vspace{-15cm}
\begin{tabular}{|c||c|c||c|c|c|c|c|}
\hline
$Y_{\rm lab}$& $W_{\rm Pb}$ (GeV)& $f_{\rm Pb}$& $W_p$ (GeV)& $f_{\rm incoh}$& $f_{\rm coh}$ & $f_{\rm incoh}$ + $f_{\rm coh}$ & $f_{\rm incoh}$ + $f_{\rm coh} (R_A = 1)$\\
\hline
-5.125 & 3570 & 0.464 $\times 10^{-12}$ &   13.31 &      6.23 &   0.0     &  6.23    &  7.16    \\
-4.875 & 3150 & 0.577 $\times 10^{-10}$ &   15.08 &      5.77 &  0.52 $\times 10^{-21}$ &  5.77    &  6.91    \\
-4.625 & 2780 & 0.930 $\times 10^{-8}$ &   17.09 &      6.37 &  0.22 $\times 10^{-10}$ &  6.37    &  6.66    \\
-4.375 & 2450 & 0.725 $\times 10^{-6}$ &   19.37 &      6.83 &  0.51 $\times 10^{-5}$ &  6.83    &  6.41    \\
-4.125 & 2160 & 0.262 $\times 10^{-4}$ &   21.95 &      7.11 &  0.39 $\times 10^{-2}$ &  7.11    &  6.16    \\
-3.875 & 1910 & 0.481 $\times 10^{-3}$ &   24.87 &      7.21 &  0.15     &  7.36    &  6.03    \\
-3.625 & 1690 & 0.500 $\times 10^{-2}$ &   28.18 &      7.17 &   1.2     &  8.35    &  6.60    \\
-3.375 & 1490 & 0.326 $\times 10^{-1}$ &   31.93 &      7.03 &   3.9     &  10.9    &  8.38    \\
-3.125 & 1310 & 0.146     &   36.18 &      6.75 &   7.5     &  14.2    &  10.9    \\
-2.875 & 1160 & 0.482     &   41.00 &      6.29 &   11     &  16.8    &  13.2    \\
-2.625 & 1020 &  1.26     &   46.46 &      5.64 &   12     &  17.8    &  14.7    \\
-2.375 &  902     &  2.71     &   52.64 &      4.97 &   12     &  17.4    &  15.5    \\
-2.125 &  796     &  5.03     &   59.65 &      4.36 &   12     &  16.3    &  15.6    \\
-1.875 &  703     &  8.33     &   67.60 &      3.81 &   11     &  14.9    &  15.4    \\
-1.625 &  620     &  12.6     &   76.60 &      3.35 &   10     &  13.4    &  14.8    \\
-1.375 &  547     &  17.7     &   86.80 &      2.96 &   9.1     &  12.1    &  14.1    \\
-1.125 &  483     &  23.5     &   98.35 &      2.64 &   8.2     &  10.9    &  13.3    \\
-0.875 &  426     &  29.9     &  111.45 &      2.36 &   7.4     &  9.77    &  12.4    \\
-0.625 &  376     &  36.7     &  126.29 &      2.11 &   6.7     &  8.79    &  11.5    \\
-0.375 &  332     &  43.8     &  143.10 &      1.89 &   6.0     &  7.91    &  10.6    \\
-0.125 &  293     &  51.2     &  162.16 &      1.69 &   5.4     &  7.09    &  9.74    \\
 0.125 &  258     &  58.6     &  183.75 &      1.51 &   4.8     &  6.33    &  8.83    \\
 0.375 &  228     &  66.2     &  208.21 &      1.33 &   4.3     &  5.61    &  7.94    \\
 0.625 &  201     &  73.9     &  235.94 &      1.17 &   3.8     &  4.93    &  7.06    \\
 0.875 &  178     &  81.6     &  267.35 &      1.02 &   3.3     &  4.29    &  6.21    \\
 1.125 &  157     &  89.4     &  302.95 &      0.88 &   2.8     &  3.68    &  5.37    \\
 1.375 &  138     &  97.2     &  343.29 &      0.74 &   2.4     &  3.11    &  4.58    \\
 1.625 &  122     &  105     &  389.00 &      0.62 &   1.9     &  2.57    &  3.82    \\
 1.875 &  108     &  113     &  440.79 &      0.51 &   1.6     &  2.08    &  3.11    \\
 2.125 &  95.1     &  121     &  499.48 &      0.40 &   1.2     &  1.63    &  2.46    \\
 2.375 &  83.9     &  128     &  565.99 &      0.31 &  0.93     &  1.24    &  1.88    \\
 2.625 &  74.0     &  136     &  641.35 &      0.23 &  0.67     & 0.903    &  1.38    \\
 2.875 &  65.3     &  144     &  726.74 &      0.17 &  0.46     & 0.630    & 0.971    \\
 3.125 &  57.7     &  152     &  823.50 &      0.12 &  0.30     & 0.419    & 0.650    \\
 3.375 &  50.9     &  160     &  933.15 &      0.08 &  0.19     & 0.266    & 0.415    \\
 3.625 &  44.9     &  167     & 1057.40 &      0.05 &  0.11     & 0.161    & 0.253    \\
 3.875 &  39.6     &  175     & 1198.19 &      0.03 &  0.60 $\times 10^{-1}$ & 0.939 $\times 10^{-1}$& 0.149    \\
 4.125 &  35.0     &  183     & 1357.73 &      0.02 &  0.32 $\times 10^{-1}$ & 0.535 $\times 10^{-1}$& 0.852 $\times 10^{-1}$\\
 4.375 &  30.9     &  191     & 1538.51 &      0.01 &  0.16 $\times 10^{-1}$ & 0.300 $\times 10^{-1}$& 0.480 $\times 10^{-1}$\\
 4.625 &  27.2     &  199     & 1743.36 &      0.01 &  0.80 $\times 10^{-2}$ & 0.165 $\times 10^{-1}$& 0.267 $\times 10^{-1}$\\
 4.875 &  24.0     &  206     & 1975.48 &      0.01 &  0.38 $\times 10^{-2}$ & 0.892 $\times 10^{-2}$& 0.145 $\times 10^{-1}$\\
 5.125 &  21.2     &  214     & 2238.52 &      0.00 &  0.18 $\times 10^{-2}$ & 0.459 $\times 10^{-2}$& 0.749 $\times 10^{-2}$\\
\hline
\end{tabular}
}
\caption{The effective photon flux $f_{\rm Pb}$ and $f_{\rm incoh},f_{\rm coh}$ radiated in the case of $\Upsilon$ production at proton-nucleon collision energy $\sqrt{s_{pN}}=5.02$ TeV by the lead ion and by the proton beam.}
\label{Tab_3}
\end{table}

 \begin{table}[]
\centering
\vspace{-2cm}
\scalebox{.9}{
\vspace{-15cm}
\begin{tabular}{|c||c|c||c|c|c|c|c|}
\hline
$Y_{\rm lab}$& $W_{\rm Pb}$ (GeV)& $f_{\rm Pb}$& $W_p$ (GeV)& $f_{\rm incoh}$& $f_{\rm coh}$ & $f_{\rm incoh}$ + $f_{\rm coh}$ & $f_{\rm incoh}$ + $f_{\rm coh} (R_A = 1)$\\
\hline
-5.125 & 2040 & 0.102 $\times 10^{-3}$ &    7.62 &      5.68 &  0.18 $\times 10^{-1}$ &  5.70    &  4.16    \\
-4.875 & 1800 & 0.145 $\times 10^{-2}$ &    8.63 &      5.50 &  0.29     &  5.79    &  4.06    \\
-4.625 & 1590 & 0.121 $\times 10^{-1}$ &    9.78 &      5.21 &   1.3     &  6.54    &  4.49    \\
-4.375 & 1400 & 0.662 $\times 10^{-1}$ &   11.08 &      4.80 &   3.1     &  7.90    &  5.44    \\
-4.125 & 1240 & 0.257     &   12.56 &      4.10 &   4.5     &  8.63    &  6.41    \\
-3.875 & 1090 & 0.759     &   14.23 &      3.31 &   5.0     &  8.30    &  7.03    \\
-3.625 &  964     &  1.81     &   16.12 &      2.55 &   4.6     &  7.17    &  7.21    \\
-3.375 &  851     &  3.63     &   18.27 &      1.93 &   3.9     &  5.80    &  7.04    \\
-3.125 &  751     &  6.38     &   20.70 &      1.46 &   3.1     &  4.58    &  6.65    \\
-2.875 &  663     &  10.1     &   23.46 &      1.12 &   2.5     &  3.58    &  6.13    \\
-2.625 &  585     &  14.7     &   26.58 &      0.87 &   1.9     &  2.80    &  5.56    \\
-2.375 &  516     &  20.2     &   30.12 &      0.69 &   1.5     &  2.22    &  4.98    \\
-2.125 &  455     &  26.2     &   34.13 &      0.56 &   1.2     &  1.78    &  4.42    \\
-1.875 &  402     &  32.8     &   38.68 &      0.47 &   1.0     &  1.46    &  3.89    \\
-1.625 &  355     &  39.8     &   43.83 &      0.39 &  0.82     &  1.21    &  3.40    \\
-1.375 &  313     &  47.0     &   49.66 &      0.34 &  0.68     &  1.02    &  2.95    \\
-1.125 &  276     &  54.4     &   56.27 &      0.29 &  0.57     & 0.860    &  2.54    \\
-0.875 &  244     &  61.9     &   63.77 &      0.25 &  0.48     & 0.733    &  2.18    \\
-0.625 &  215     &  69.5     &   72.26 &      0.22 &  0.40     & 0.626    &  1.85    \\
-0.375 &  190     &  77.2     &   81.88 &      0.20 &  0.34     & 0.535    &  1.57    \\
-0.125 &  168     &  84.9     &   92.78 &      0.17 &  0.28     & 0.456    &  1.32    \\
 0.125 &  148     &  92.7     &  105.13 &      0.15 &  0.24     & 0.389    &  1.11    \\
 0.375 &  130     &  100     &  119.13 &      0.13 &  0.20     & 0.330    & 0.922    \\
 0.625 &  115     &  108     &  134.99 &      0.12 &  0.16     & 0.279    & 0.764    \\
 0.875 &  102     &  116     &  152.97 &      0.10 &  0.13     & 0.235    & 0.630    \\
 1.125 &  89.7     &  124     &  173.34 &      0.09 &  0.11     & 0.197    & 0.517    \\
 1.375 &  79.1     &  132     &  196.42 &      0.08 &  0.86 $\times 10^{-1}$ & 0.164    & 0.423    \\
 1.625 &  69.8     &  139     &  222.57 &      0.07 &  0.69 $\times 10^{-1}$ & 0.136    & 0.344    \\
 1.875 &  61.6     &  147     &  252.20 &      0.06 &  0.54 $\times 10^{-1}$ & 0.112    & 0.279    \\
 2.125 &  54.4     &  155     &  285.78 &      0.05 &  0.42 $\times 10^{-1}$ & 0.918 $\times 10^{-1}$& 0.225    \\
 2.375 &  48.0     &  163     &  323.83 &      0.04 &  0.33 $\times 10^{-1}$ & 0.748 $\times 10^{-1}$& 0.181    \\
 2.625 &  42.4     &  170     &  366.95 &      0.04 &  0.25 $\times 10^{-1}$ & 0.605 $\times 10^{-1}$& 0.145    \\
 2.875 &  37.4     &  178     &  415.81 &      0.03 &  0.19 $\times 10^{-1}$ & 0.486 $\times 10^{-1}$& 0.115    \\
 3.125 &  33.0     &  186     &  471.18 &      0.02 &  0.14 $\times 10^{-1}$ & 0.388 $\times 10^{-1}$& 0.911 $\times 10^{-1}$\\
 3.375 &  29.1     &  194     &  533.91 &      0.02 &  0.11 $\times 10^{-1}$ & 0.307 $\times 10^{-1}$& 0.717 $\times 10^{-1}$\\
 3.625 &  25.7     &  202     &  605.00 &      0.02 &  0.78 $\times 10^{-2}$ & 0.242 $\times 10^{-1}$& 0.562 $\times 10^{-1}$\\
 3.875 &  22.7     &  209     &  685.56 &      0.01 &  0.56 $\times 10^{-2}$ & 0.190 $\times 10^{-1}$& 0.437 $\times 10^{-1}$\\
 4.125 &  20.0     &  217     &  776.84 &      0.01 &  0.40 $\times 10^{-2}$ & 0.148 $\times 10^{-1}$& 0.339 $\times 10^{-1}$\\
 4.375 &  17.7     &  225     &  880.27 &      0.01 &  0.29 $\times 10^{-2}$ & 0.114 $\times 10^{-1}$& 0.262 $\times 10^{-1}$\\
 4.625 &  15.6     &  233     &  997.48 &      0.01 &  0.20 $\times 10^{-2}$ & 0.884 $\times 10^{-2}$& 0.202 $\times 10^{-1}$\\
 4.875 &  13.8     &  240     & 1130.29 &      0.01 &  0.14 $\times 10^{-2}$ & 0.673 $\times 10^{-2}$& 0.154 $\times 10^{-1}$\\
 5.125 &  12.1     &  248     & 1280.79 &      0.00 &  0.96 $\times 10^{-3}$ & 0.505 $\times 10^{-2}$& 0.116 $\times 10^{-1}$\\
\hline
\end{tabular}
}
\caption{The effective photon flux $f_{\rm Pb}$ and $f_{\rm incoh},f_{\rm coh}$ radiated in the case of $J/\psi$ production at proton-nucleon collision energy $\sqrt{s_{pN}}=5.02$ TeV by the lead ion and by the proton beam.}
\label{Tab_4}
\end{table}

\begin{table}[]
\centering
\vspace{-2cm}
\scalebox{.9}{
\vspace{-15cm}
\begin{tabular}{|c||c|c||c|c|c|c|c|}
\hline
$Y_{\rm lab}$& $W_{\rm Pb}$ (GeV)& $f_{\rm Pb}$& $W_p$ (GeV)& $f_{\rm incoh}$& $f_{\rm coh}$ & $f_{\rm incoh}$ + $f_{\rm coh}$ & $f_{\rm incoh}$ + $f_{\rm coh} (R_A = 1)$\\
\hline
-5.125 & 4550 & 0.548 $\times 10^{-8}$ &   16.97 &      7.14 &  0.90 $\times 10^{-11}$ &  7.14    &  7.52    \\
-4.875 & 4010 & 0.482 $\times 10^{-6}$ &   19.23 &      7.70 &  0.35 $\times 10^{-5}$ &  7.70    &  7.27    \\
-4.625 & 3540 & 0.190 $\times 10^{-4}$ &   21.79 &      8.07 &  0.33 $\times 10^{-2}$ &  8.07    &  7.02    \\
-4.375 & 3130 & 0.374 $\times 10^{-3}$ &   24.69 &      8.22 &  0.15     &  8.37    &  6.88    \\
-4.125 & 2760 & 0.410 $\times 10^{-2}$ &   27.98 &      8.22 &   1.2     &  9.47    &  7.49    \\
-3.875 & 2430 & 0.279 $\times 10^{-1}$ &   31.70 &      8.11 &   4.2     &  12.3    &  9.51    \\
-3.625 & 2150 & 0.129     &   35.92 &      7.85 &   8.4     &  16.3    &  12.4    \\
-3.375 & 1900 & 0.437     &   40.70 &      7.37 &   12     &  19.5    &  15.2    \\
-3.125 & 1670 &  1.16     &   46.12 &      6.65 &   14     &  20.8    &  17.2    \\
-2.875 & 1480 &  2.54     &   52.26 &      5.92 &   15     &  20.7    &  18.3    \\
-2.625 & 1300 &  4.78     &   59.22 &      5.23 &   14     &  19.5    &  18.7    \\
-2.375 & 1150 &  7.99     &   67.11 &      4.61 &   13     &  18.0    &  18.5    \\
-2.125 & 1010 &  12.2     &   76.04 &      4.08 &   12     &  16.4    &  18.0    \\
-1.875 &  896     &  17.2     &   86.17 &      3.65 &   11     &  14.9    &  17.4    \\
-1.625 &  790     &  23.0     &   97.64 &      3.28 &   10     &  13.6    &  16.6    \\
-1.375 &  697     &  29.3     &  110.64 &      2.97 &   9.4     &  12.4    &  15.7    \\
-1.125 &  615     &  36.1     &  125.38 &      2.69 &   8.6     &  11.3    &  14.8    \\
-0.875 &  543     &  43.2     &  142.07 &      2.45 &   7.9     &  10.3    &  13.9    \\
-0.625 &  479     &  50.5     &  160.99 &      2.23 &   7.2     &  9.43    &  12.9    \\
-0.375 &  423     &  58.0     &  182.42 &      2.03 &   6.6     &  8.60    &  12.0    \\
-0.125 &  373     &  65.6     &  206.71 &      1.84 &   6.0     &  7.82    &  11.1    \\
 0.125 &  329     &  73.2     &  234.23 &      1.67 &   5.4     &  7.07    &  10.1    \\
 0.375 &  291     &  80.9     &  265.42 &      1.50 &   4.9     &  6.37    &  9.21    \\
 0.625 &  257     &  88.7     &  300.76 &      1.34 &   4.4     &  5.69    &  8.31    \\
 0.875 &  226     &  96.5     &  340.81 &      1.19 &   3.9     &  5.04    &  7.43    \\
 1.125 &  200     &  104     &  386.19 &      1.04 &   3.4     &  4.42    &  6.57    \\
 1.375 &  176     &  112     &  437.61 &      0.91 &   2.9     &  3.83    &  5.73    \\
 1.625 &  156     &  120     &  495.87 &      0.78 &   2.5     &  3.27    &  4.93    \\
 1.875 &  137     &  128     &  561.90 &      0.65 &   2.1     &  2.75    &  4.17    \\
 2.125 &  121     &  135     &  636.72 &      0.54 &   1.7     &  2.26    &  3.45    \\
 2.375 &  107     &  143     &  721.49 &      0.44 &   1.4     &  1.81    &  2.79    \\
 2.625 &  94.4     &  151     &  817.56 &      0.35 &   1.1     &  1.41    &  2.19    \\
 2.875 &  83.3     &  159     &  926.42 &      0.27 &  0.79     &  1.06    &  1.66    \\
 3.125 &  73.5     &  167     & 1049.77 &      0.20 &  0.57     & 0.766    &  1.20    \\
 3.375 &  64.9     &  174     & 1189.54 &      0.14 &  0.39     & 0.530    & 0.837    \\
 3.625 &  57.2     &  182     & 1347.93 &      0.10 &  0.25     & 0.348    & 0.555    \\
 3.875 &  50.5     &  190     & 1527.40 &      0.07 &  0.15     & 0.219    & 0.351    \\
 4.125 &  44.6     &  198     & 1730.77 &      0.04 &  0.87 $\times 10^{-1}$ & 0.132    & 0.212    \\
 4.375 &  39.3     &  206     & 1961.22 &      0.03 &  0.48 $\times 10^{-1}$ & 0.766 $\times 10^{-1}$& 0.124    \\
 4.625 &  34.7     &  213     & 2222.35 &      0.02 &  0.25 $\times 10^{-1}$ & 0.435 $\times 10^{-1}$& 0.710 $\times 10^{-1}$\\
 4.875 &  30.6     &  221     & 2518.26 &      0.01 &  0.13 $\times 10^{-1}$ & 0.244 $\times 10^{-1}$& 0.400 $\times 10^{-1}$\\
 5.125 &  27.0     &  229     & 2853.56 &      0.01 &  0.62 $\times 10^{-2}$ & 0.135 $\times 10^{-1}$& 0.222 $\times 10^{-1}$\\
\hline
\end{tabular}
}
\caption{The effective photon flux $f_{\rm Pb}$ and $f_{\rm incoh},f_{\rm coh}$ radiated in the case of $\Upsilon$ production at proton-nucleon collision energy $\sqrt{s_{pN}}=8.16$ TeV by the lead ion and by the proton beam.}
\label{Tab_1}
\end{table}

\begin{table}[]
\centering
\vspace{-2cm}
\scalebox{.9}{
\vspace{-15cm}
\begin{tabular}{|c||c|c||c|c|c|c|c|}
\hline
$Y_{\rm lab}$& $W_{\rm Pb}$ (GeV)& $f_{\rm Pb}$& $W_p$ (GeV)& $f_{\rm incoh}$& $f_{\rm coh}$ & $f_{\rm incoh}$ + $f_{\rm coh}$ & $f_{\rm incoh}$ + $f_{\rm coh} (R_A = 1)$\\
\hline
-5.125 & 2600 & 0.102 $\times 10^{-1}$ &    9.71 &      5.80 &   1.4     &  7.19    &  4.94    \\
-4.875 & 2300 & 0.577 $\times 10^{-1}$ &   11.00 &      5.37 &   3.3     &  8.70    &  5.99    \\
-4.625 & 2030 & 0.230     &   12.47 &      4.63 &   5.0     &  9.62    &  7.11    \\
-4.375 & 1790 & 0.695     &   14.13 &      3.76 &   5.6     &  9.36    &  7.86    \\
-4.125 & 1580 &  1.68     &   16.01 &      2.91 &   5.2     &  8.15    &  8.11    \\
-3.875 & 1390 &  3.43     &   18.14 &      2.21 &   4.4     &  6.64    &  7.97    \\
-3.625 & 1230 &  6.10     &   20.55 &      1.68 &   3.6     &  5.27    &  7.56    \\
-3.375 & 1080 &  9.73     &   23.29 &      1.30 &   2.8     &  4.13    &  7.01    \\
-3.125 &  957     &  14.3     &   26.39 &      1.01 &   2.2     &  3.25    &  6.39    \\
-2.875 &  845     &  19.7     &   29.90 &      0.80 &   1.8     &  2.57    &  5.76    \\
-2.625 &  746     &  25.7     &   33.89 &      0.66 &   1.4     &  2.09    &  5.14    \\
-2.375 &  658     &  32.2     &   38.40 &      0.55 &   1.2     &  1.72    &  4.55    \\
-2.125 &  581     &  39.2     &   43.51 &      0.46 &  0.97     &  1.43    &  4.00    \\
-1.875 &  512     &  46.4     &   49.30 &      0.40 &  0.81     &  1.21    &  3.49    \\
-1.625 &  452     &  53.7     &   55.87 &      0.34 &  0.68     &  1.03    &  3.03    \\
-1.375 &  399     &  61.3     &   63.31 &      0.30 &  0.58     & 0.880    &  2.62    \\
-1.125 &  352     &  68.9     &   71.74 &      0.27 &  0.49     & 0.758    &  2.25    \\
-0.875 &  311     &  76.6     &   81.29 &      0.24 &  0.42     & 0.653    &  1.92    \\
-0.625 &  274     &  84.3     &   92.11 &      0.21 &  0.35     & 0.562    &  1.63    \\
-0.375 &  242     &  92.0     &  104.37 &      0.19 &  0.30     & 0.484    &  1.38    \\
-0.125 &  214     &  99.8     &  118.27 &      0.17 &  0.25     & 0.415    &  1.16    \\
 0.125 &  188     &  108     &  134.02 &      0.15 &  0.21     & 0.356    & 0.974    \\
 0.375 &  166     &  115     &  151.86 &      0.13 &  0.17     & 0.303    & 0.814    \\
 0.625 &  147     &  123     &  172.08 &      0.12 &  0.14     & 0.258    & 0.678    \\
 0.875 &  130     &  131     &  195.00 &      0.10 &  0.12     & 0.218    & 0.564    \\
 1.125 &  114     &  139     &  220.96 &      0.09 &  0.95 $\times 10^{-1}$ & 0.184    & 0.467    \\
 1.375 &  101     &  146     &  250.38 &      0.08 &  0.77 $\times 10^{-1}$ & 0.155    & 0.387    \\
 1.625 &  89.0     &  154     &  283.72 &      0.07 &  0.62 $\times 10^{-1}$ & 0.130    & 0.319    \\
 1.875 &  78.6     &  162     &  321.50 &      0.06 &  0.49 $\times 10^{-1}$ & 0.109    & 0.263    \\
 2.125 &  69.3     &  170     &  364.30 &      0.05 &  0.39 $\times 10^{-1}$ & 0.904 $\times 10^{-1}$& 0.217    \\
 2.375 &  61.2     &  178     &  412.81 &      0.04 &  0.31 $\times 10^{-1}$ & 0.750 $\times 10^{-1}$& 0.178    \\
 2.625 &  54.0     &  185     &  467.77 &      0.04 &  0.24 $\times 10^{-1}$ & 0.619 $\times 10^{-1}$& 0.145    \\
 2.875 &  47.7     &  193     &  530.06 &      0.03 &  0.19 $\times 10^{-1}$ & 0.508 $\times 10^{-1}$& 0.119    \\
 3.125 &  42.1     &  201     &  600.63 &      0.03 &  0.14 $\times 10^{-1}$ & 0.416 $\times 10^{-1}$& 0.963 $\times 10^{-1}$\\
 3.375 &  37.1     &  209     &  680.61 &      0.02 &  0.11 $\times 10^{-1}$ & 0.338 $\times 10^{-1}$& 0.778 $\times 10^{-1}$\\
 3.625 &  32.8     &  216     &  771.23 &      0.02 &  0.83 $\times 10^{-2}$ & 0.273 $\times 10^{-1}$& 0.627 $\times 10^{-1}$\\
 3.875 &  28.9     &  224     &  873.92 &      0.02 &  0.62 $\times 10^{-2}$ & 0.219 $\times 10^{-1}$& 0.501 $\times 10^{-1}$\\
 4.125 &  25.5     &  232     &  990.28 &      0.01 &  0.46 $\times 10^{-2}$ & 0.174 $\times 10^{-1}$& 0.398 $\times 10^{-1}$\\
 4.375 &  22.5     &  240     & 1122.13 &      0.01 &  0.33 $\times 10^{-2}$ & 0.138 $\times 10^{-1}$& 0.315 $\times 10^{-1}$\\
 4.625 &  19.9     &  248     & 1271.54 &      0.01 &  0.24 $\times 10^{-2}$ & 0.108 $\times 10^{-1}$& 0.248 $\times 10^{-1}$\\
 4.875 &  17.5     &  255     & 1440.85 &      0.01 &  0.17 $\times 10^{-2}$ & 0.852 $\times 10^{-2}$& 0.195 $\times 10^{-1}$\\
 5.125 &  15.5     &  263     & 1632.69 &      0.01 &  0.12 $\times 10^{-2}$ & 0.663 $\times 10^{-2}$& 0.152 $\times 10^{-1}$\\
\hline
\end{tabular}
}
\caption{The effective photon flux $f_{\rm Pb}$ and $f_{\rm incoh},f_{\rm coh}$ radiated in the case of $J/\psi$ production at proton-nucleon collision energy $\sqrt{s_{pN}}=8.16$ TeV by the lead ion and by the proton beam.}
\label{Tab_2}
\end{table}

\section*{Acknowledgements}

We thank Petja Paakkinen for providing the numerical values for the EPPS16 gluon modification factor at the scales $\mu^2 = m_c^2$ and $\mu^2 = m_b^2$ used in this work. CAF is supported by the Helsinki Institute of Physics core funding project QCD-THEORY. SPJ is supported by a Royal Society University Research Fellowship (Grant URF/R1/201268). The work of TT is supported by the STFC Consolidated Grant ST/T000988/1. This work is also supported in part by the STFC Grant ST/T001011/1.

\newpage
\bibliographystyle{unsrt}
\bibliography{references.bib}
\endthebibliography

\end{document}